\documentclass[11pt,hyper]{JHEP3}
\usepackage{epsfig}

\def\lsim{\mathrel{\rlap{\lower4pt\hbox{\hskip1pt$\sim$}}
    \raise1pt\hbox{$<$}}}         
\def\gsim{\mathrel{\rlap{\lower4pt\hbox{\hskip1pt$\sim$}}
    \raise1pt\hbox{$>$}}}         
\newcommand{\anl}{\\[1ex]}

\newcommand{\ds}{{\sffamily DarkSUSY}}

\newcommand{\ft}{\tilde{f}}

\newcommand{\snu}{\tilde{\nu}}

\newcommand{\gt}{\tilde{g}}

\newcommand{\into}{\rightarrow}
        
\newcommand{\bea}{\begin{eqnarray}}
\newcommand{\eea}{\end{eqnarray}}
\newcommand{\be}{\begin{equation}}
\newcommand{\ee}{\end{equation}}
\newcommand{\ba}{\begin{array}}
\newcommand{\ea}{\end{array}}

\title{Accurate relic densities with neutralino, chargino and sfermion 
coannihilations in mSUGRA}

\author{Joakim Edsj\"o, Mia Schelke\\ Department of Physics, AlbaNova, Stockholm University, SE-106 91 Stockholm, Sweden\\ 
E-mail: \email{edsjo@physto.se, schelke@physto.se}}
\author{Piero Ullio\\ SISSA, via Beirut 4, I-34014 Trieste, Italy\\
  E-mail: \email{ullio@sissa.it}}
\author{Paolo Gondolo\\Department of Physics, Case Western Reserve University, 
10900 Euclid Ave., Cleveland OH 44106-7079, USA\\
   E-mail: \email{pxg26@po.cwru.edu}}

\abstract{Neutralinos arise as natural dark matter candidates in many 
supersymmetric extensions of the Standard Model. We present a novel 
calculation of the neutralino relic abundance in which we include all so 
called coannihilation processes between neutralinos, charginos and sfermions,
and, at the same time, we apply the state of the art technique to trace the 
freeze-out of a species in the early Universe. 
As a first application, we discuss here results valid in the mSUGRA framework;
we enlight general trends as well as perform a detailed study of the
neutralino relic densities in the mSUGRA parameter space.
The emerging picture is fair agreement with previous analyses in the same 
framework, however we have the power to discuss it in many more details
than previously done. E.g., we find that the cosmological bound on the neutralino mass is pushed up to $\sim 565$ GeV in the stau coannihilation region and to $\sim 1500$ GeV in the chargino coannihilation region.
}

\keywords{supersymmetry, dark matter}
\preprint{\hepph{0301106}\\
NSF-ITP-03-03\\
CWRU-P1-03}

\begin{document}

\section{Introduction}

The latest years will be remembered in the history of Science as those
that marked the entrance into the era of precision Cosmology. A number
of experiments have been pinning down the values of cosmological
parameters to a level of precision hardly foreseeable just a decade ago,
with perspectives from upcoming measurements even more spectacular.
Most remarkably, experiments with different focus, as well
as exploiting complementary techniques, have all collected data which 
point 
to one single overall-consistent picture, a ``concordance'' 
model~\cite{triang} in which the Universe is flat with about 30\% of 
its present average energy density in matter and about 70\%
in some form with negative pressure (a cosmological constant or dark 
energy). More precisely, from the combined analysis of the latest data
on the cosmic microwave background and large scale galaxy surveys,
the cold dark matter (CDM) and baryonic contributions have been recently 
estimated~\cite{hubble} to be
$\Omega_{CDM}h^2 = 0.115 \pm 0.009$ and $\Omega_{b}h^2 = 0.022 \pm 0.002$,
respectively (here $\Omega$ is 
the ratio between mean density and critical density $\rho_c=1.879 \times
10^{-29}  h^2 $ g/cm$^3$, and $h$ is 
the Hubble constant in units of 100 km s$^{-1}$ Mpc$^{-1}$; the analysis 
in
\cite{hubble} gives $h = 0.665 \pm 0.047$).

Among the ideas which have been put forward to account for
the CDM term, the most natural solution is probably the scheme in which 
CDM appears as a thermal leftover from the early 
Universe: in this context, stable weakly interacting massive particles 
(WIMPs) are ideal CDM candidates, as their thermal relic abundance 
is naturally of the order of the measured one.
Given the accuracy of the current and future 
measurements of $\Omega_{CDM}$, it is useful to have
an equally accurate calculation of the
relic abundance of WIMPs.
This paper
continues our program of accurate relic density computations.

Since the late seventies,
when the idea of WIMP dark matter was formulated~\cite{leewei,gunn,steig},
the computation of the WIMP relic abundance has been constantly refined.
One important step was to recognize the role of coannihilation 
effects~\cite{BGS,GS}: if in the particle physics theory one 
considers, a stable WIMP appears together with a slightly heavier 
particle into which it can transform, when computing the present
density of the lightest particle one needs to retrace the thermal
history of both particles simultaneously.
Such an effect is common even for the most popular WIMP dark matter 
candidate, the lightest supersymmetric particle (LSP) in supersymmetric 
extensions of the particle physics Standard Model.
When the LSP is the lightest neutralino,
the mass splitting between 
the LSP and the next-to-lightest supersymmetric particle may in some cases 
be
small. Coannihilations are then important.
Their inclusion in the computation of the neutralino relic density 
has been the subject of numerous 
studies, which vary in the degrees of refinement of
the method implemented to compute the relic abundance
and in the selection of which particles to include in the coannihilating 
set.

The novel calculation that we present in this paper is included in 
a new extended version of the \ds\ package \cite{ds}, which will be 
publicly available in the near future. \ds\ now allows for the most 
generic coannihilation effect in the
framework of the minimal supersymmetric extension of the Standard Model
(MSSM), at the same time applying the state of the art technique to 
trace the freeze-out of a species in the early 
Universe~\cite{GondoloGelmini}. In fact, on 
one side, we fully solve the density evolution equation (which determines the 
evolution of the number density of neutralinos) numerically, including all 
possible resonance and threshold effects, and avoiding approximations
in the thermally averaged cross sections (such as the expansions in terms 
of the relative velocity that is often applied). 
On the other side, extending the work of two of us~\cite{eg-coann} 
who first applied such a technique to the case of 
coannihilations with charginos and neutralinos,
we include here the 
possibility of having coannihilations with all sfermions as well. 
As a result, assuming masses, widths and couplings of particles in the MSSM
are given with an adequate precision, 
we provide here a tool to compute neutralino relic 
abundances with an estimated precision of 1\% or better.

Compared to other recent calculations, we believe this is the most 
accurate calculation available at present. The standard lore so far
has been to calculate the thermal average of the
annihilation cross section by
expanding to first power in temperature over mass and implementing
an approximate solution to the evolution equation which estimates the 
freeze out temperature without fully solving the equation
(see, e.g., Kolb and Turner~\cite{KT}). Sometimes this is refined by
including resonances and threshold corrections~\cite{GS}.
Among recent studies, this approach is taken in e.g.~Refs.~\cite{BDD,GLP}.
Other refinements include, e.g., solving the density evolution equation
numerically but still using an approximation to thermal effects in the
cross section~\cite{Ellisstau1,Ellisstau2,Ellisstop,Ellislast,Ellislargetg},
or calculating the thermal average accurately but using an
approximate solution to the density 
equation~\cite{micromega,bbb,roszkowski}. At the same time, only in a 
few of the quoted papers the full set of initial states has been included.
As already mentioned, the present calculation includes all initial states,
performs an accurate thermal average and gives a very accurate solution to 
the evolution equation. Though the inclusion of initial state sfermions in 
the \ds\ package is a new feature introduced in the present work, other 
groups~\cite{GLP2,BKKZ,BKKG} have earlier introduced some sfermion coannihilations 
in an interface with the old \ds\ version.

The \ds\ package has been written in a very
general and flexible format, so that it can be used for any theory 
embedded in the MSSM. As a first example, we will present here results 
valid in the minimal supergravity (mSUGRA) framework. The mSUGRA framework
 is the 
framework considered in most of the previous analyses.
Although rather restrictive in the way parameters are set, it is sufficient
to enlight most coannihilation effects which can emerge in the MSSM.

The outline of this paper is the following: in the next section we 
discuss the supersymmetric model we work in; in section \ref{sec:coann} 
we review the framework in which we calculate the relic density 
including coannihilations. 
In section 4 we examine the effects of coannihilation in detail, 
 stressing the physical insights of the results.

\section{Supersymmetric model}

The particle physics model implemented in the \ds\ package is a minimal 
supersymmetric extension of the Standard Model, built with $N=1$ generator 
of supersymmetry and containing the smallest possible number of fields. 
We restrict ourselves to the case in which the LSP is
the lightest neutralino, i.e.~the lightest of the four mass eigenstates
obtained from the superposition of the supersymmetric partners of the 
neutral gauge and Higgs bosons. The supersymmetric part
of the spectrum contains also two chargino mass eigenstates, the gluino
and the scalar superpartners of leptons and quarks. 
The Higgs sector needs two Higgs doublets, which, after electroweak 
symmetry breaking, give five scalar fields, denoted as $H^\pm, H^0_1, H^0_2$ 
and $H^0_3$ (for a more detailed description of the model and of our 
conventions see~\cite{bg}).\footnote{Apart from sfermion coannihilations, 
another major improvement in the \ds\ version that we use for this work is 
the inclusion of one-loop corrections to the Higgs widths, with formulas taken 
from Refs.~\cite{DSZ,Spira}.}

As needed in the relic density calculation, we perform 
the computation of all two-body final state cross sections at tree level 
for all initial states involving neutralinos, charginos, sleptons and 
squarks. We do not neglect mass 
terms for fermions in final states (as sometimes done in the literature), nor 
implement any expansion to first order in relative momentum of the incoming 
particles (as most often done in the past).\footnote{In the current
 version of \ds\ we do not include processes 
with the gluino in the initial state or any flavour changing processes
in the sfermion coannihilations; these might be included in future work 
but we do not expect any result in the present analysis to be affected.} 
A list of all processes included is given in 
Appendix \ref{sec:processes}. 

We compute analytic expressions for 
the amplitudes 
using standard Feynman rules with generic 
expressions for the vertex couplings (e.g.\ $i g^L_{\phi f f'} P_L + i
g^R_{\phi f f'} P_R$ for a vertex involving two fermions and a scalar).
The amplitudes for all the different types of Feynman diagrams are
obtained with the help of symbolic manipulation programs. 
For initial states involving neutralinos and 
charginos,
helicity amplitudes for each 
type of diagram are computed analytically with {\sc Reduce}~\cite{reduce}
(this is described in \cite{eg-coann}\footnote{Compared to the
  analysis in \cite{eg-coann}, we have here corrected a sign error in the
chargino -- (down-type) fermion -- (up-type) sfermion vertex with clashing arrows.
Consequently, the discrepancy between DarkSUSY and
micrOMEGAs (model B in \cite{micromega}) has now disappeared.}). 
For the diagrams including sfermions in the initial state,
{\sc Form} \cite{form} is used to analytically calculate the
amplitudes squared. The output from {\sc Form} 
is then converted into Fortran with a {\sc
Perl} script. All of these analytic formulas 
can be converted into a compact 
form, but we do not consider it useful to reproduce those expressions here. 

The actual values of the MSSM vertex couplings are introduced during 
the numerical
calculation. We recomputed all vertex couplings from the MSSM lagrangian, and
checked them against
standard literature (e.g.~\cite{HK,couplings}; 
\ds\ also contains couplings  which do
not appear in the literature, namely those involving 
generic intergenerational mixing in the sfermion sector). 

When giving explicit examples of computations of the neutralino
relic abundance we restrict ourselves in this paper to mSUGRA models.
Under the assumption of universality at the grand unification scale, 
the mSUGRA action has five free parameters,
$m_{1/2},  m_0,  sign(\mu),  A_0$ and $\tan\beta$. The parameters
$m_{1/2}$, $m_0$ and  $A_0$ are the GUT unification values 
of the soft supersymmetry breaking fermionic mass parameters,
scalar mass parameters and trilinear scalar coupling parameters, respectively
(of the trilinear couplings, only $A_t$, $A_b$ and $A_\tau$ differ from zero).
The absolute value of the Higgs superfield parameter $\mu$ follows from 
electroweak symmetry breaking, but its sign is free. Finally,
in the Higgs sector, $\tan\beta$ denotes the ratio, $v_2/v_1$, of the vacuum
expectation values of the two neutral components of the SU(2) Higgs doublets. 
Our convention on the sign of $\mu$ is that $\mu$ appears with a minus 
sign in the superpotential (i.e.~following the convention 
used in e.g.~\cite{HK}), 
while the definition of sign 
and dimension for the $A_0$'s are such that $A_t$ appears in the stop 
mass squared matrix as off-diagonal elements of the form 
$(A_t-\mu\cot\beta) m_t$ (and analogously for $A_b$ and $A_\tau$).

GUT scale values of the soft breaking parameters, as well as of the gauge 
and Yukawa couplings, have to be evolved down to the weak scale. To do 
that we make use of the ISASUGRA RGE package in the ISAJET 7.64 
software~\cite{isajet}\footnote{The ISASUGRA output is rather unstable in regions where 
the convergence is slow (e.g.~in the focus point region where the lightest chargino 
is nearly degenerate with the lightest neutralino); to improve the stability, we 
have converted ISASUGRA to double precision and increased the requirements on the 
convergence.}. The interface
to \ds\ is such that the whole SUSY spectrum of masses and mixings as given 
in the output at the weak scale in ISASUGRA is given as input in the 
\ds\ MSSM spectrum. 
Regarding Standard Model masses and couplings, we have implemented
1-loop renormalization group equations. We fix the top pole mass at
174.3 GeV, which is the central value stated by the Particle Data Group 
2002 \cite{pdg02}. For the $c$ and $b$ quarks we input the running
quark masses $m_q(m_q)$ and choose the values of 1.26 GeV 
and 4.2 GeV respectively~\cite{pdg02,KS}.
To a good approximation, the RGE scale at
which (co)annihilations occur is twice the LSP mass,
and therefore we evaluate the gauge and Yukawa couplings
at this scale. 

We have chosen to use the ISASUGRA RGE code since it is a widely used program that is kind of a standard in the field. However, there are other RGE codes on the market (e.g.\ SOFTSUSY \cite{softsusy}, SPHENO \cite{spheno} and SUSPECT \cite{suspect}), all of which solve the RGEs with a different level of sophistication. We do not want to
go into the details of these different packages here, but instead refer the interested reader to a recent comparison in Ref.\ \cite{rge-comp}. The main message we want to convey here is that different RGE codes give slightly different results. E.g.\ the sparticle masses typically differ by a few GeV between the different codes. 

When scanning over the parameter space, we compare with accelerator 
constraints, implementing limits recommended
by the Particle Data Group 2002 (PDG) \cite{pdg02}. 
We have not included limits which are still preliminary as of this writing
(most important among the preliminary bounds is the limit 
of about 104 GeV on the chargino mass;
we have nevertheless indicated its effect in some of the figures). 
For the Higgs's 
$H^0_2$ and $H^0_3$ we have implemented the $\tan\beta$ dependent 
mass limits as provided by the PDG in the most conservative setup. This 
is particularly important for the mass of $H^0_2$ since for a 
substantial $\tan\beta$ interval this constraint is less restrictive 
than the lower mass limit (114.3 GeV) on the Standard Model Higgs.
In this paper, which focusses on coannihilations, we do not emphasize the role
of the 
$b \rightarrow s \gamma$ constraint (we refer the interested reader 
to recent dedicated analyses in the same mSUGRA context, e.g.,
Refs.~\cite{bsgamma}).

\section{Relic density calculation}
\label{sec:coann}

The importance of coannihilations in the computation of the density 
of a relic particle was recognized by Binetruy, Girardi and 
Salati~\cite{BGS}, and independently by 
Griest and Seckel \cite{GS}. In Edsj\"o and Gondolo
\cite{eg-coann}, this was further analysed and put in a form that
allows for an accurate treatment of coannihilations in the same basic
framework as without coannihilations. We do not repeat every step of
that calculation here. Instead, we only briefly review the results
we need for this paper. For more details, we refer the reader to
Ref.~\cite{eg-coann}.

\subsection{The density evolution equation and thermal averaging}

We consider the coannihilation of $N$ species of particles ($\chi_i$, 
$i=1,\ldots,N$) with masses
$m_i$ and internal degrees of freedom (statistical weights) $g_i$ (see
Appendix~\ref{appdof} for conventions on degrees of freedom adopted in this paper).
We order
the masses in increasing order, $m_1 \le m_2 \le \cdots \le m_N$,
 and use $m_1$ and $m_\chi$
interchangeably for the mass of the lightest particle.  If the lightest 
particle is stable and the others decay into it, 
instead of considering the thermal history of each
particle separately, we can follow the evolution of the sum of the number
densities. The problem is then formulated in terms of the 
density evolution equation \cite{GS,eg-coann}
\begin{equation} \label{eq:Boltzmann2}
  \frac{dn}{dt} =
  -3Hn - \langle \sigma_{\rm{eff}} v \rangle 
  \left( n^2 - n_{\rm{eq}}^2 \right)
\end{equation}
where $\langle \sigma_{\rm eff} v \rangle$ is the effective 
thermally-averaged annihilation cross section, $H$ is the Hubble parameter, and $n$ is the total number
density summed over all coannihilating particles. 
The effective thermally-averaged
annihilation cross section is
\begin{equation} \label{eq:sigmaveff}
  \langle \sigma_{\rm{eff}} v \rangle = 
  \frac{A}{n_{\rm{eq}}^2} \, ,
\end{equation}
where $n_{\rm eq}$ is the equilibrium number density, which in the 
Maxwell-Boltzmann approximation (which is a good approximation for our case) 
reads
\begin{equation} \label{eq:neq}
  n_{\rm eq} = 
  \frac{T}{2\pi^2} \sum_i g_i m_{i}^2
  \, K_{2} \!\left( \frac{m_{i}}{T}\right);
\end{equation}
and $A$ is the annihilation
rate per unit volume at temperature $T$
\begin{equation}
\label{eq:Apeff}
  A = \frac{g_1^2 T}{4 \pi^4} \int_{0}^\infty dp_{\rm eff}
  p^2_{\rm eff} W_{\rm eff} \, K_{1} \!
  \left( \frac{\sqrt{s}}{T}\right) .
\end{equation}
Here $K_{i}(x)$, ($i=1,2$), are the modified Bessel functions of the second
kind of order $i$; $s= 4p_{\rm{eff}}^2 + 4m_\chi^2$ is the usual Mandelstam
variable giving the center-of-mass energy squared for the $\chi_1\chi_1$
system; $p_{\rm eff}$ is the center-of-mass momentum for the $\chi_1\chi_1$
system; and $W_{\rm eff}$ is the effective annihilation rate obtained by
summing over all annihilation and coannihilation channels,
\begin{equation} \label{eq:weff}
  W_{\rm{eff}} = \sum_{ij}\frac{p_{ij}}{p_{11}}
  \frac{g_ig_j}{g_1^2} W_{ij} = 
  \sum_{ij} \sqrt{\frac{[s-(m_{i}-m_{j})^2][s-(m_{i}+m_{j})^2]}
  {s(s-4m_1^2)}} \frac{g_ig_j}{g_1^2} W_{ij}.
\end{equation}
For the coannihilation of particles $i$
and $j$, 
$W_{ij}$ is the annihilation rate per unit volume and unit time given by 
\footnote{The quantity $w_{ij}$ in Ref.\ \protect\cite{SWO} is
  $W_{ij}/4$, and is therefore one-fourth of the annihilation rate per unit
  volume and unit time.}
\begin{equation} \label{eq:Wijcross}
  W_{ij} = 4 p_{ij} \sqrt{s} \sigma_{ij} = 4 \sigma_{ij} \sqrt{(p_i
\cdot p_j)^2 - m_i^2 m_j^2} = 4 E_{i} E_{j} \sigma_{ij} v_{ij} .
\end{equation}
where
\begin{equation}
  p_{ij} =
\frac{\left[s-(m_i+m_j)^2\right]^{1/2}
\left[s-(m_i-m_j)^2\right]^{1/2}}{2\sqrt{s}}
\end{equation}
is the common magnitude of the 3-momentum of particle $i$ and $j$ in the
center-of-mass frame of the pair $\chi_i\chi_j$.
Defining  $W_{ij}(s) = 0 $ for $s \le (m_i+m_j)^2$, 
the radicand in Eq.~(\ref{eq:weff}) is  never negative.
The effective thermally averaged annihilation cross section can then be 
written as
\begin{equation} \label{eq:sigmavefffin2}
  \langle \sigma_{\rm{eff}}v \rangle = \frac{\int_0^\infty
  dp_{\rm{eff}} p_{\rm{eff}}^2 W_{\rm{eff}} \,K_1 \!\left(
  \frac{\sqrt{s}}{T} \right) } { m_1^4 T \left[ \sum_i \frac{g_i}{g_1}
  \frac{m_i^2}{m_1^2} \,K_2 \!\left(\frac{m_i}{T}\right) \right]^2}.
\end{equation}
This expression is very similar to the one in the case of no coannihilations
given in Gondolo and Gelmini~\cite{GondoloGelmini} (and correctly 
reduces to it in the absence of coannihilations). The only differences 
are in the denominator and in the replacement of the annihilation rate 
with the effective annihilation rate. The key feature in 
Eq.~(\ref{eq:sigmavefffin2}) is the definition of an effective 
annihilation rate independent of temperature, with $p_{\rm eff}$ as 
integration variable. This gives a remarkable calculational advantage,
as $W_{\rm eff}$ can be tabulated in advance, before taking the thermal 
average and solving the density evolution equation.

The steps we implement to compute the relic density are the following. We first
rephrase the density evolution equation (\ref{eq:Boltzmann2}) as an equation
for $Y=n/s$, with $s$ being the entropy density. Then we tabulate $W_{\rm eff}$
including thresholds, resonances and coannihilations, and spline it.  As a
third step we solve the density evolution equation starting from the boundary
condition which sets particles in equilibrium at the temperature $T=
m_{\chi}/2$ (a specially-devised implicit method is used in the
numerical integration of the density equation).  This implies actually a double
integration since at each temperature step we need to calculate the thermal
average $\langle \sigma_{\rm{eff}}v \rangle$ (this integration is relatively
fast as we use the $W_{\rm eff}$ tabulation). The integration of the density
equation is performed until $Y$ has reached a constant value (the freeze out
value) which we finally convert into the value of the relic abundance.

As stated earlier, we estimate our calculation of the relic density to be accurate to about 1\%. We base this estimate on the precision with which we do the tabulation of the effective annihilation rate $W_{\rm eff}$ and on the precision with which we perform the numerical integrations. For example, we explicitly make sure that resonances and thresholds are tabulated and integrated with such a precision that the end result is accurate to at least 1\%. 
One should keep in mind that this is the accuracy of the calculation of 
the neutralino relic density starting from given values of masses, widths 
and couplings of MSSM particles. In the mSUGRA framework considered here, the accuracy of the sparticle masses for a given set of input parameters is less than 1\% (see e.g.\ \cite{rge-comp} for a comparison of different RGE-codes) and the main uncertainty in 
evaluating the neutralino relic density in mSUGRA thus comes from the RGE code ISASUGRA.

\subsection{A few examples of coannihilation effects}
\label{omegaex}

For illustrative purposes only, we rewrite Eq.~(\ref{eq:sigmavefffin2}) 
in the form:
\begin{equation} \label{eq:sigmavefffin3}
  \langle \sigma_{\rm{eff}}v \rangle = \int_0^\infty
  dp_{\rm{eff}} \frac{W_{\rm{eff}}(p_{\rm{eff}})}{4\,E_{\rm{eff}}^2} 
  \kappa(p_{\rm{eff}},T)\;,
\end{equation}
where $E_{\rm{eff}} = \sqrt{p_{\rm{eff}}^2 + m_{\chi}^2}$
is the energy per particle in the center of mass for 
$\chi_1\chi_1$ annihilations. 
The term we factor out, $W_{\rm{eff}}/4{E^2_{\rm{eff}}}$, 
can be thought of as
an effective $\sigma v$ term (compare with Eq.~(\ref{eq:Wijcross})).
In the $p_{\rm{eff}}
\rightarrow 0$ limit, $W_{\rm{eff}}/4{E^2_{\rm{eff}}}$ reduces to 
the $\chi_1$$\chi_1$ annihilation rate
at zero temperature, which is the relevant quantity in indirect dark 
matter detection searches. The term $\kappa$ we introduced in 
Eq.~(\ref{eq:sigmavefffin3}) contains the Boltzmann factor and the 
phase-space integrand term and can be regarded as a weight function,
at the temperature $T$, that selects which range of $p_{\rm{eff}}$ 
is important in the thermal average. As the phase-space integrand term 
dominates at small $p_{\rm{eff}}$ and makes $\kappa$ go to 0 in the 
$p_{\rm{eff}} \rightarrow 0$ limit, $\kappa$ shows a peak at an 
intermediate $p_{\rm{eff}}$ and then rapidly decreases due to the 
Boltzmann suppression; the position and height of the peak depends 
on the temperature considered and on the particles involved.

\FIGURE[t]{
  \centerline{\epsfig{file=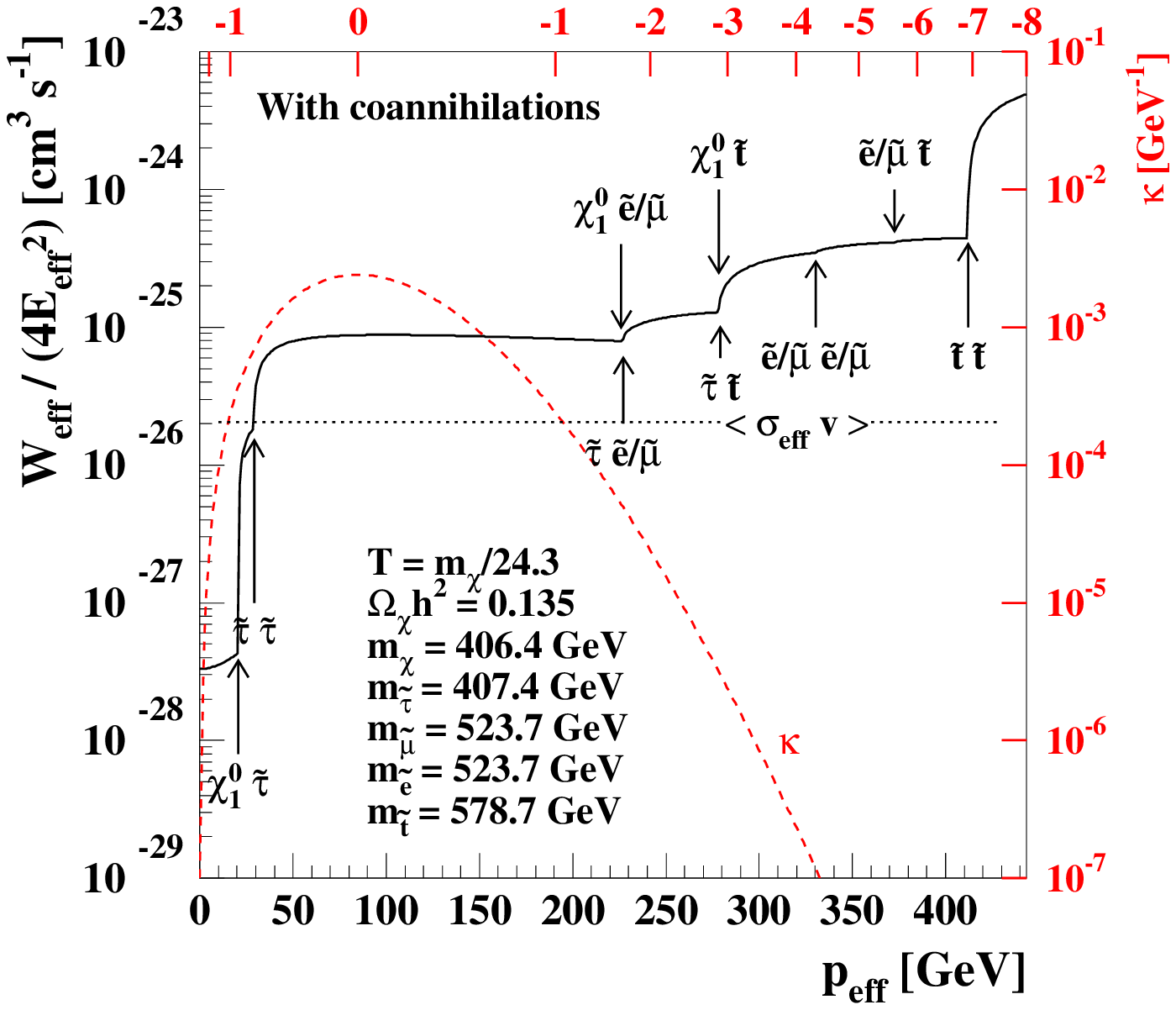,width=0.49\textwidth}
  \epsfig{file=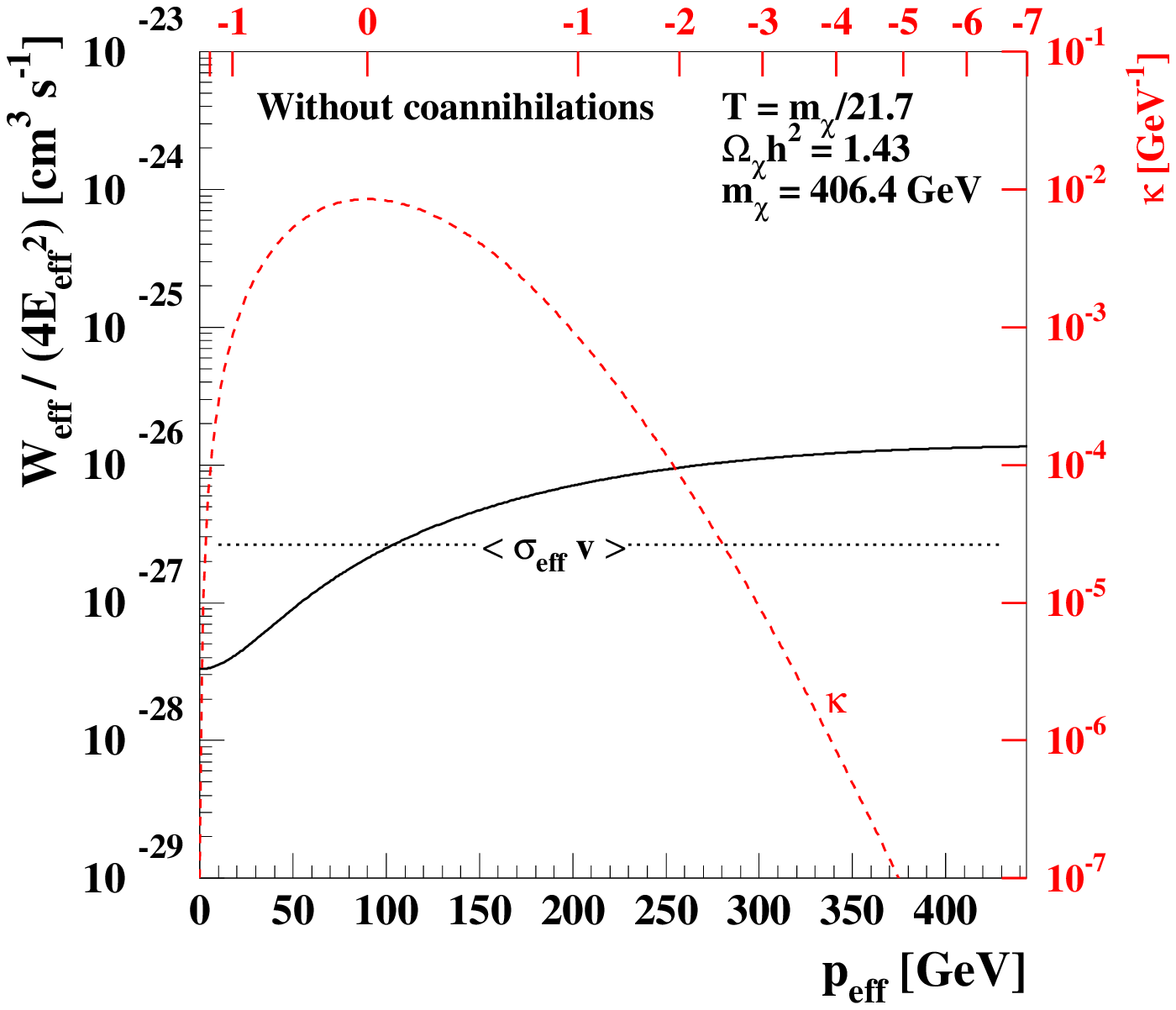,width=0.49\textwidth}}
  \caption{The effective annihilation cross section a) with coannihilations and
    b) without coannihilations for model A (specified in
    Table~\ref{tab:examples} in Appendix \ref{appexmodels}). The solid line
    shows the effective annihilation cross section $W_{\rm eff}/4 E_{\rm
      eff}^2$ as a function of momentum $p_{\rm eff}$, while the dashed line
    shows the thermal weight factor $\kappa(p_{\rm eff},T)$. The
    thermally-averaged annihilation cross section is the integral over $p_{\rm
      eff}$ of the product of the two.  Note that when including
    coannihilations, not only new thresholds appear, but the freeze-out
    temperature is also changing, meaning that we sample a different region of
    the annihilation cross section. For this model, the relic density with
    coannihilations is $\Omega_{\chi,\, \rm coann}h^2=0.135$ and that without
    is $\Omega_{\chi,\, \rm no~coann}h^2 = 1.43$.}
  \label{fig:weff-sf-coann}}

We are now ready to show some examples of coannihilation effects.  As already
mentioned, the examples we display have the lightest neutralino as the LSP and
are in the mSUGRA framework. In Fig.~\ref{fig:weff-sf-coann}a we consider a
case in which the neutralino, with mass of about 400~GeV, is nearly mass
degenerate with the lightest stau. The lightest selectron, the lightest 
smuon and the lightest stop are relatively close in mass as well. (To fully
specify the example models we present, the model 
parameters and some properties are
given in Table \ref{tab:examples} in Appendix \ref{appexmodels}. The model in
Fig.~\ref{fig:weff-sf-coann} is model A in that table.) The solid curve shows
$W_{\rm{eff}}/ 4 E^2_{\rm{eff}}$, and one can nicely see coannihilations
appearing as thresholds at $\sqrt{s}$ equal to the sum of the masses of the
coannihilating particles (just as final state thresholds do). As usually
happens when considering coannihilation effects with neutralinos as the LSP,
the $\chi^0_1$-$\chi^0_1$ contribution to $W_{\rm{eff}}$ is small compared with
the one provided by the coannihilating particles. The role of coannihilating
particles can be quantified better with a look at the function $\kappa$ (dashed
curve, in units of GeV$^{-1}$, and with relative scale shown on the right-hand
side of the figure). The factor $\kappa$ 
is plotted at the freeze out
temperature, defined as the temperature at which the abundance of the relic
species is 50\% higher than the equilibrium value\footnote{This is given here
  for illustrative purposes only; it is never actually exploited in the full
  computation since we solve the density evolution equation numerically.}, in
this case $T=m_{\chi}/24.3$.  On the top of the panel, the tick mark labelled
'0' indicates the position of the momentum $p_{\rm{eff}}^{\rm{max}}$
corresponding to the maximum of $\kappa$, while the other tick marks indicate
the momenta $p_{\rm{eff}}^{(n)}$ at which $\kappa$ is $10^{-n}$ of its maximum
value, 
$\kappa(p_{\rm{eff}}^{(n)})/\kappa(p_{\rm{eff}}^{\rm{max}}) = 10^{-n}$.
The tick marks provide a 
visual guide to the interval in $p_{\rm{eff}}$ which is relevant in the thermal
averaging. The integral of the product of $W_{\rm{eff}}/ 4 E^2_{\rm{eff}}$ and
$\kappa$ gives $\langle
\sigma_{\rm{eff}}v \rangle$ thermally averaged at the freeze out temperature
(shown in the figure as a horizontal dotted line). This is the quantity which is 
sufficient to get a rough indication of the neutralino relic abundance through 
the rule of thumb~\cite{jkg}
$\Omega_\chi h^2 \simeq 10^{-27}$~cm$^3$~s$^{-1}/\langle \sigma_{\rm{eff}}v \rangle$.

In Fig.~\ref{fig:weff-sf-coann}b we consider the same model but ignore
coannihilation effects. One can see that $W_{\rm{eff}}/ 4 E^2_{\rm{eff}}$ is
now, on average, much smaller, and therefore one can expect the relic abundance
to be higher. This is indeed the case, with a shift from $\Omega_\chi h^2 =
0.135$ including coannihilations (left panel) to $ \Omega_\chi h^2 = 1.43$ when
coannihilations are neglected (right panel).  Note, however, that the change in
$\Omega_\chi h^2$ is smaller than what one would naively expect from
comparing the solid curves in the two panels. This is due to the fact that there 
is a significant change in the freeze out temperature as well,  from
$T=m_{\chi}/24.3$ to $T=m_{\chi}/21.7$.  The weight function $\kappa$ for this
new temperature is shown in the figure, and comparing it to the one in the left
panel, one clearly sees the change in normalization (partially due to the
change in the number of degrees of freedom involved in the two cases, see the
denominator in Eq.~(\ref{eq:sigmavefffin2})) and in width (for the latter
effect, note the shift in the scale shown on the top of the figure, while the
displayed range in $p_{\rm{eff}}$ has been kept fixed). The net result is that
$\langle \sigma_{\rm{eff}}v \rangle$ at the freeze out temperature is lowered
by just about an order of magnitude, and then the increase in the relic
abundance is of the same order. From this discussion it is  evident that a
very accurate solution of the density evolution equation is needed to claim
good accuracy on the estimate of the relic density.

\FIGURE[t]{
  \centerline{\epsfig{file=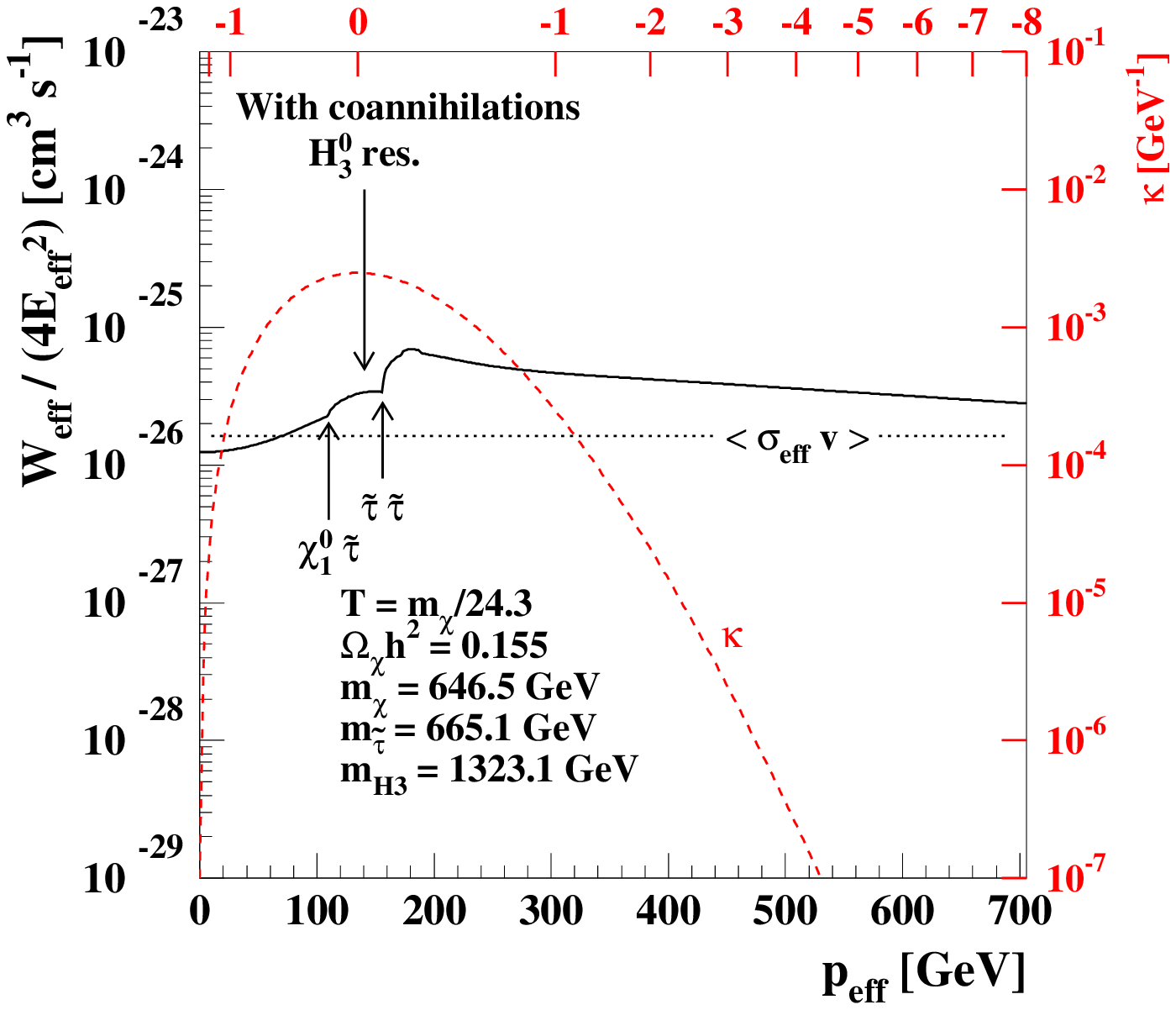,width=0.49\textwidth}
  \epsfig{file=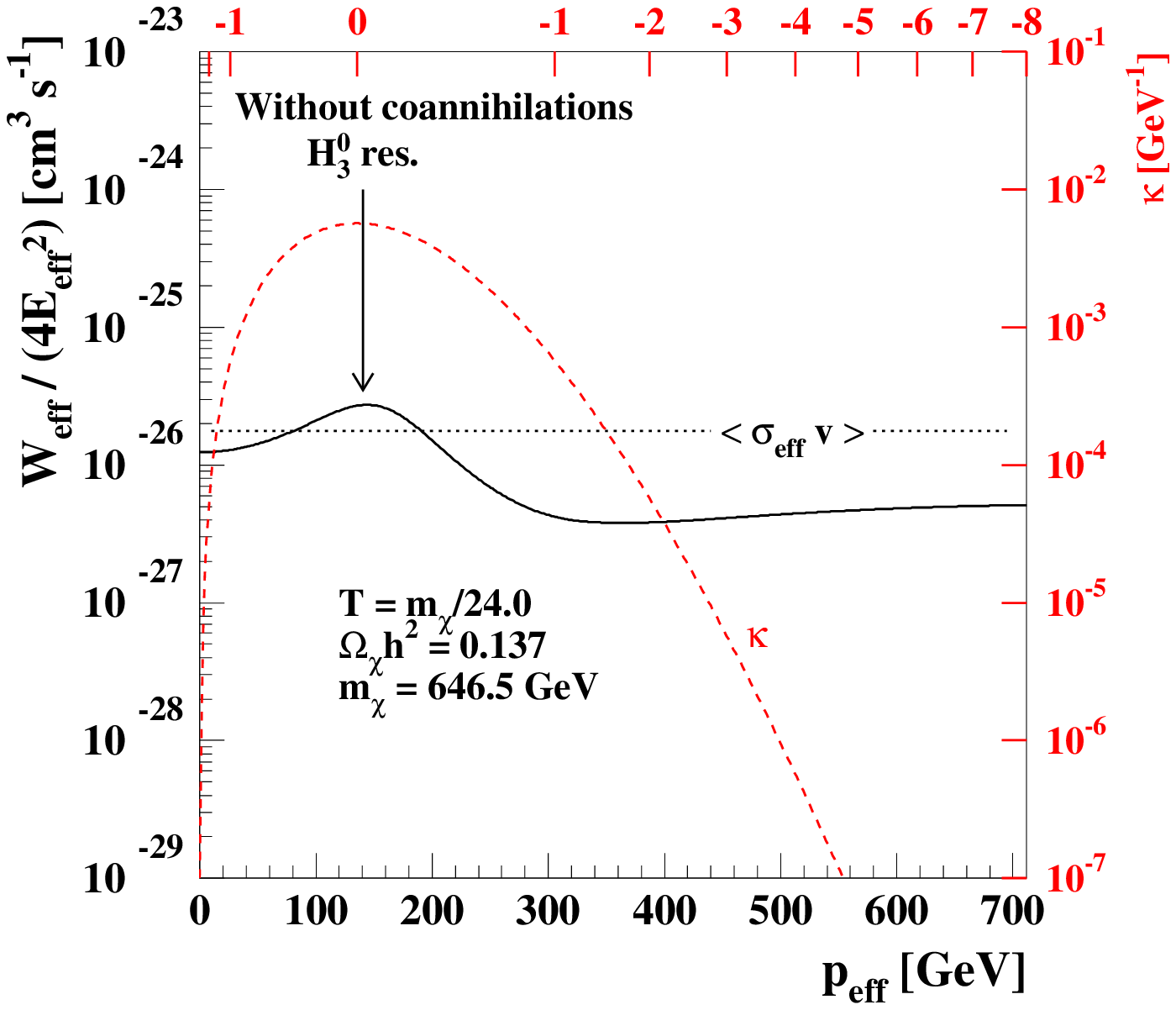,width=0.49\textwidth}}
  \caption{We here show the effective annihilation cross section 
  versus $p_{\rm eff}$ for an example (model B in Table~\ref{tab:examples} 
  in Appendix \ref{appexmodels}), where coannihilations increase the relic 
  density (in this case from 0.137 to 0.155).}
  \label{fig:weff-sf-incr}}

We mentioned that usually, when considering neutralino dark matter, 
$W_{\rm{eff}}$ increases sharply when coannihilating particles are 
included. The reason is that the coannihilating particles typically 
have non zero electric or colour charges, while the neutralino interacts only 
weakly. In the MSSM there are a few exceptions to this general trend of increasing 
$W_{\rm{eff}}$ (see ~\cite{eg-coann}), and we find one such exception in the mSUGRA 
framework as well. In Fig.~\ref{fig:weff-sf-incr}a
we show a model (model B in Table~\ref{tab:examples} in Appendix \ref{appexmodels}) 
for which the neutralino mass is slightly below half of the 
$H^0_3$ mass and $\tan\beta$ is large. The $\chi^0_1$-$\chi^0_1$ annihilation rate 
is dominated by the s-channel $H^0_3$ resonance and is quite large. (There is also 
a $H^0_1$ resonance at this mass, but it is subdominant with respect to the 
$H^0_3$ resonance). The effect
of stau coannihilating particles comes on top of that, but the total contribution
to $W_{\rm{eff}}$ is just about of the same order as from the 
$\chi^0_1$-$\chi^0_1$ term. If we now go to the case when coannihilations
are neglected, Fig.~\ref{fig:weff-sf-incr}b,
we see that the freeze out temperature remains about the same,
but still there is a shift in the normalization of the weight function due to
different numbers of degrees of freedom in the two cases (see the
denominator in Eq.~(\ref{eq:sigmavefffin2})). We find that the
net effect is a slight increase in $\langle \sigma_{\rm{eff}}v \rangle$, which
suggests that the relic abundance should be smaller if coannihilations are 
neglected. This is confirmed by the full solution of the density evolution
 equation:
the calculation that includes coannihilations yields 
$\Omega_\chi h^2=0.155$, while if stau coannihilation is neglected
one obtains $\Omega_\chi h^2=0.137$.
Therefore one should keep
in mind that although the general trend is that coannihilation effects lower 
the neutralino relic abundances, in some cases they can increase it.

\FIGURE[t]{
  \centerline{\epsfig{file=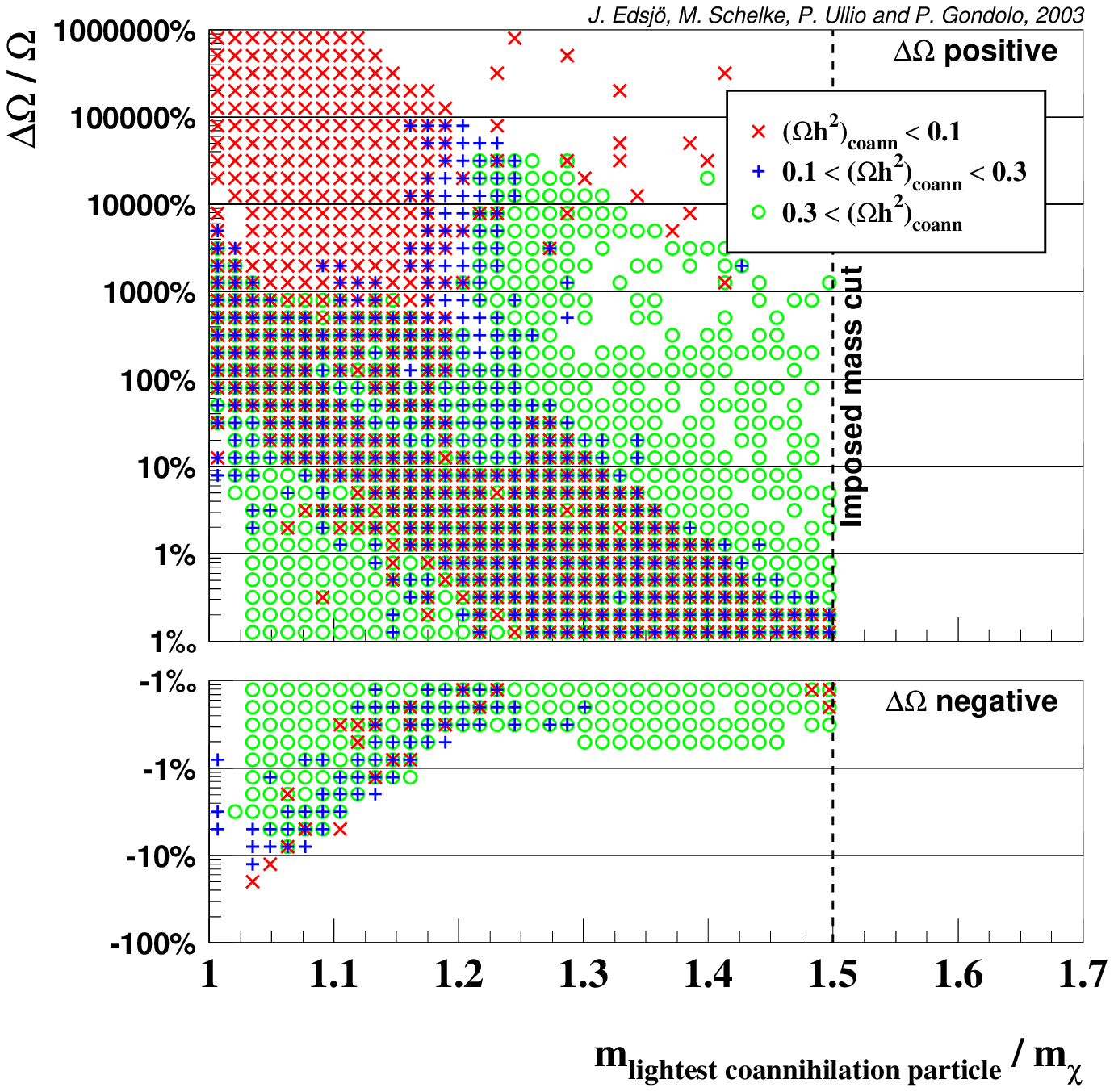,width=0.7\textwidth}}
  \caption{The relative change in relic density due to the inclusion of coannihilations,
   $\Delta \Omega / \Omega \equiv (\Omega_{\rm no~coann}-\Omega_{\rm coann})/
  \Omega_{\rm coann}$, versus the mass ratio between the lightest coannihilating particle 
  and the lightest neutralino. 
  To gain computational speed we
  avoid including unnecessary coannihilation processes in the calculation by
  imposing a cut at $1.5m_\chi$.
  The models included in the figure are from
  some general mSUGRA scans with $m_0\in[0,2500]$ GeV, $m_{1/2}\in[0,2500]$ GeV,  
  $\tan\beta\in[5,50]$,
  $A_0\in[-5000,5000]$ GeV and $sign(\mu)=\pm$. All models in
  Figs.~\ref{fig:chargino-tg30}--\ref{fig:stop-tg10-oh2diff} are also included here.
  }
  \label{fig:oh2r-dm}}

Note that due to the Boltzmann suppression of heavier particles, we do
not need to include all supersymmetric particles in the
calculation. By extensive scans  of the mSUGRA parameter space and
estimating the effects of including different particles in the
calculation, we have found that a convenient criterion to have an
accuracy of 1\% or better on the relic density in cases that are 
cosmologically interesting is to include all supersymmetric particles 
with a mass below $1.5m_\chi$. This is shown in Fig.~\ref{fig:oh2r-dm} 
where we plot the relative difference in relic density without coannihilations
and with coannihilations,
\begin{equation}
  \frac{\Delta\Omega}{\Omega} = 
  \frac{\Omega_{\chi,\, \rm {no \,coann}}-\Omega_{\chi\, \rm {coann}}}
  {\Omega_{\chi,\, \rm {coann}}} \,.
\end{equation}
versus the mass ratio of the lightest
coannihilating particle and the neutralino in our sample of models: the coding
in relic density shows that, for cosmologically interesting cases,
only mass differences up to about 40\% are important, hence the cut at the 
mass ratio 1.5 (vertical dashed line in the figure) is sufficiently
conservative. If a 1\% accuracy is required even for models that are 
cosmologically disfavoured, one would need to raise the cut
to about 1.7: the coannihilation effects one picks in this way, however,
correspond to the $\chi_1^0-\chi_1^0$ annihilation cross section being 
very suppressed and, even adding coannihilation terms, the relic 
abundance still remains $\Omega_\chi h^2 \gg 1$. 
In Fig.~\ref{fig:weff-stop} we show the effective annihilation cross
section for such an example (model C in Table~\ref{tab:examples} in 
Appendix \ref{appexmodels}). In this case stop coannihilations are important
even though the stop mass is about 50\% higher than the neutralino
mass, but still the relic density, which is $\Omega_\chi h^2=73.2$ when
coannihilations are neglected, is shifted down to just 
$\Omega_\chi h^2=19.7$. 
The criterion with selection on mass we give is the easiest to implement
in the numerical calculation, but clearly depends on the strength of the 
interactions we are dealing with; we can say that it is perfectly safe
in the scheme we are dealing with, but need not be 
valid in other schemes.

\FIGURE[t]{
  \centerline{\epsfig{file=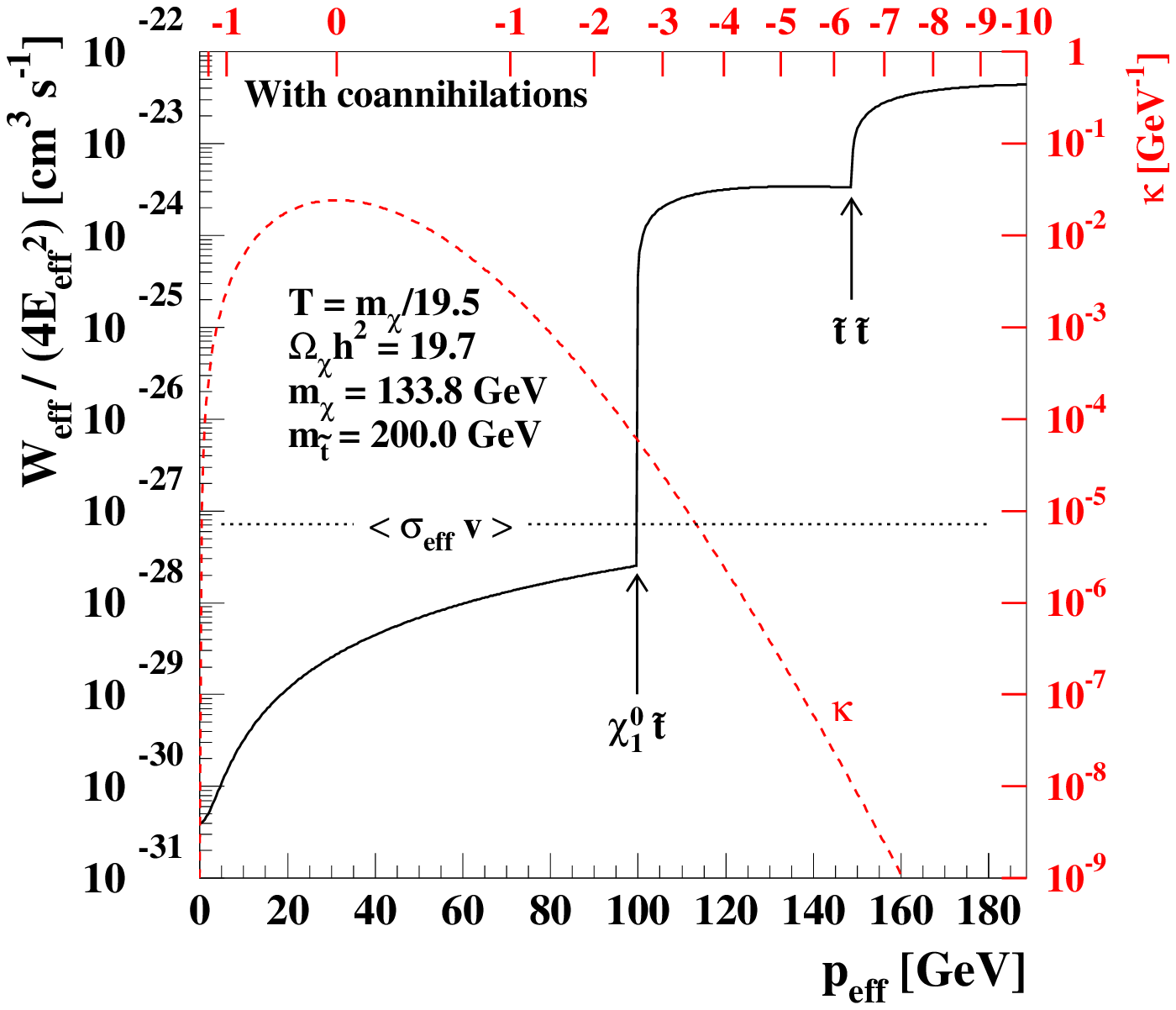,width=0.49\textwidth}}
  \caption{The effective annihilation cross section
    as a function of $p_{\rm eff}$ for an example model (model C in 
    Table~\ref{tab:examples} in Appendix \ref{appexmodels}) where stop
    coannihilations are important even though the lightest stop is
    about 50\% heavier than the lightest neutralino.}
  \label{fig:weff-stop}}

\FIGURE[t]{
  \centerline{\epsfig{file=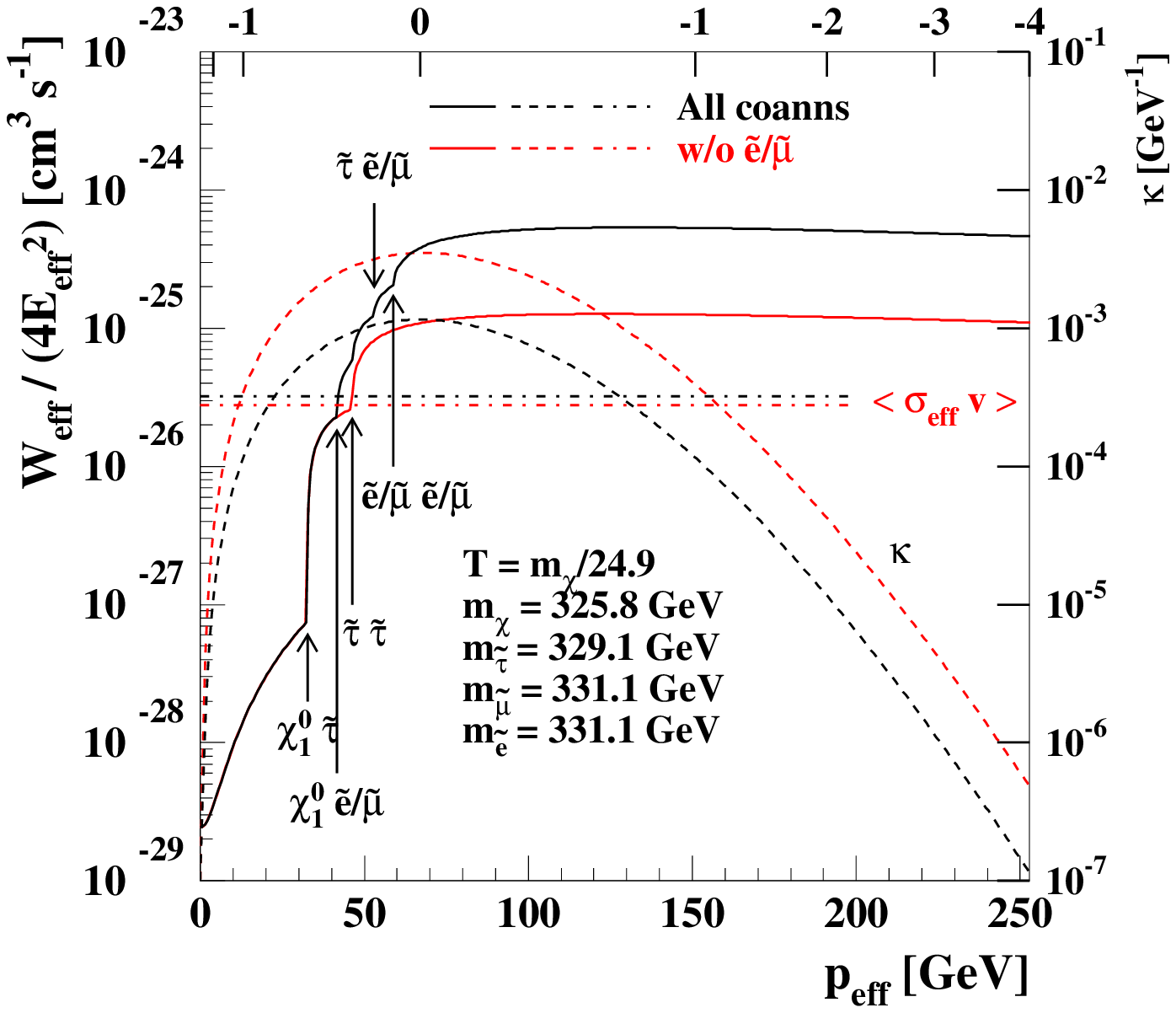,width=0.49\textwidth}}
  \caption{The effective annihilation cross section versus
    $p_{\rm eff}$ 
    for an example (model D in Table~\ref{tab:examples} in
    Appendix \ref{appexmodels}) where inclusion of all three lightest sleptons is 
    important to get the correct relic density.}
  \label{fig:weff-smuimp}}

A final check we perform is the following: sometimes in the literature only the
next-to-lightest (NLSP) sparticle has been included in the coannihilation 
calculation and one may question whether this can be considered a fair approximation. 
There is reason to believe that also heavier sparticles might change the relic LSP
density by a non-negligible amount if they are either close in mass to the NLSP or 
if they have larger coupling strength than the NLSP (as e.g.~$\chi^\pm_2$ has compared 
to $\chi^0_2$). Whether or not it is a good approximation to include only NLSP 
coannihilations should therefore be checked case by case. Even restricting to the 
cases of coannihilating particles with comparable couplings, such as staus, smuons 
and selectrons, it is not straightforward to provide a firm criterion telling when to 
include all of them or just the NLSP (i.e.~the $\tilde{\tau}$ in this case) in the 
calculation. In practice, we find regions in the mSUGRA parameter space when smuon 
and selectron contributions give corrections to the relic density of the
order of 10\% or higher. An example  is given in Fig.~\ref{fig:weff-smuimp} 
(model D in Table~\ref{tab:examples} in Appendix 
\ref{appexmodels}). In this example, the lightest 
smuon and selectron are almost degenerate in mass with both the lightest stau and
lightest neutralino. Including in the calculation the coannihilation with these three sfermions 
result in a relic density $\Omega_\chi h^2=0.109$. If instead we only include 
$\chi^0\tilde{\tau}$ and $\tilde{\tau}\tilde{\tau}$ coannihilations, the relic 
density would be 0.128, i.e.~we would be $\sim18\%$
off from the correct value.  Hence, as this example has shown, it can be
important to include the coannihilations of more than one particle. To be on
the safe side, we always include {\it all} particles with masses up to
$1.5m_\chi$.

\section{The role of coannihilations in the mSUGRA framework}

The mSUGRA framework~\cite{mSUGRA} has been extensively discussed in the 
literature. We summarize here the main features that
are relevant for understanding the results of the relic abundance computation.
In line with most previous analyses, we sample the 5-dimensional mSUGRA
parameter space choosing a few values of $\tan \beta$ and $A_0$, and slices
along the $m_{1/2}, m_0$ planes for both $sign(\mu)$. 

Consider first the case $A_0=0$. Two regimes with cosmologically interesting 
relic abundances have been identified: The region $m_0 \lsim m_{1/2}$ 
and the region $m_0 \gg m_{1/2}$. In the first region, 
the lightest neutralino is a 
quasi pure bino with mass set essentially by $m_{1/2}$ alone; 
the parameter $m_0$ sets
the sfermion mass scale, with the slepton sector lighter than the squark 
sector and with the lightest stau always being the lightest sfermion,
possibly lighter than the lightest neutralino if $m_0 \ll m_{1/2}$.
In the second regime, the relevant region is a narrow band, sometimes dubbed 
the ``focus point'' region \cite{focus-point},
close to the region where there is no radiative 
electro-weak symmetry breaking: in such a band the parameter $\mu$ 
is driven to small values and forces a mixing between the gaugino and 
Higgsino
sectors. As a consequence the lightest neutralino may contain a 
large Higgsino fraction and the next-to-lightest supersymmetric particle 
is a chargino. Large values of $A_0$ can introduce a third regime in the 
intermediate $m_0$ 
range: off-diagonal entries in the stop mass matrix can become sufficiently 
large to drive the lightest stop to masses smaller than the mass of the 
lightest stau or even the mass of the lightest neutralino.
On top of these generic trends, details are sensitive to the value of 
$\tan \beta$ and the $sign(\mu)$.

In each of the regimes above, the mass splitting between the LSP,
which we require to be the lightest neutralino, and the next-to-lightest
SUSY particle can be small enough for coannihilation effects to become
important. We label the three regions as the slepton coannihilation region,
the chargino coannihilation region and the stop coannihilation region,
and describe each of them separately.

\subsection{Slepton coannihilations}

\FIGURE[t]{
\centerline{\epsfig{file=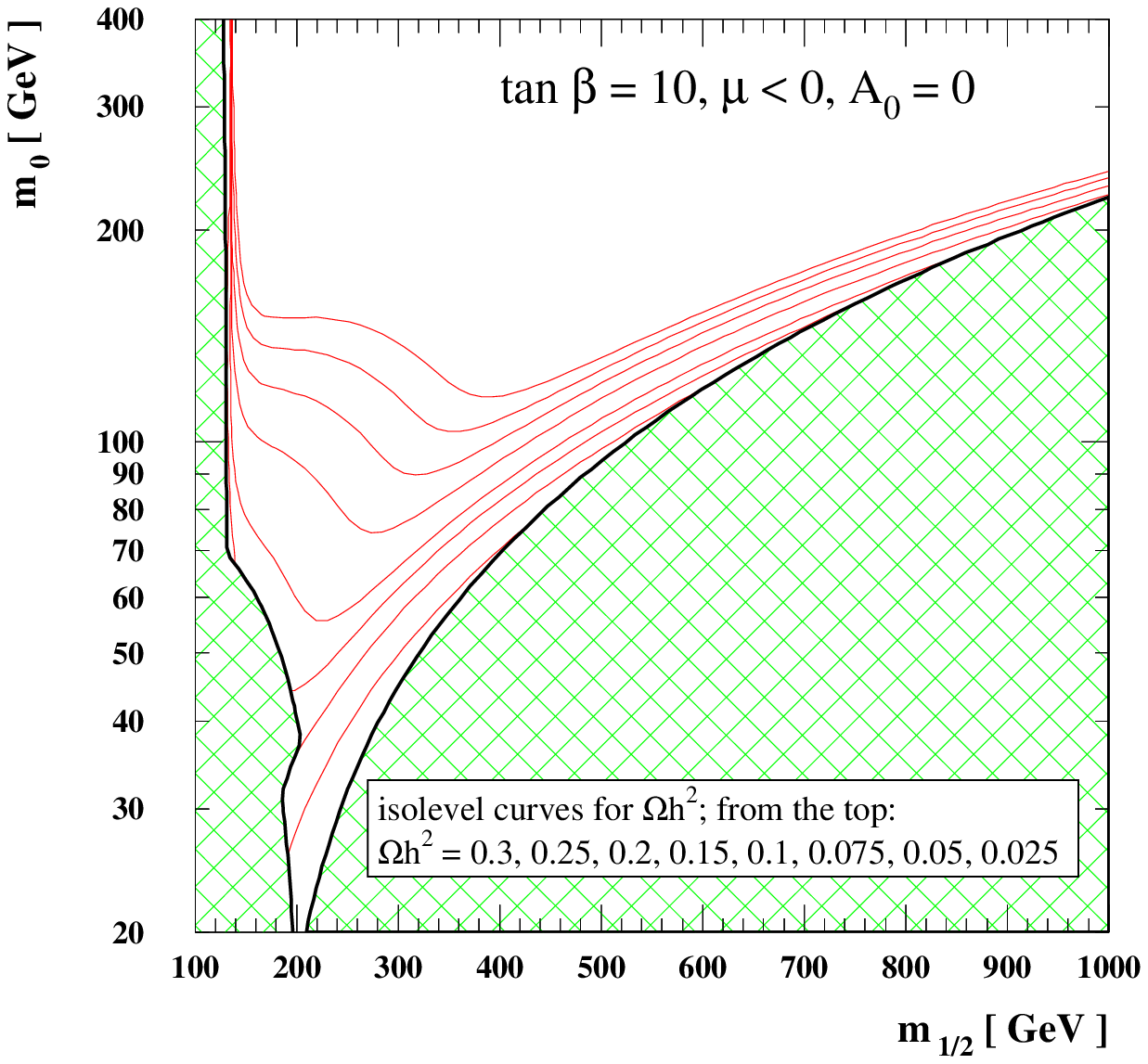,width=0.49\textwidth}
\epsfig{file=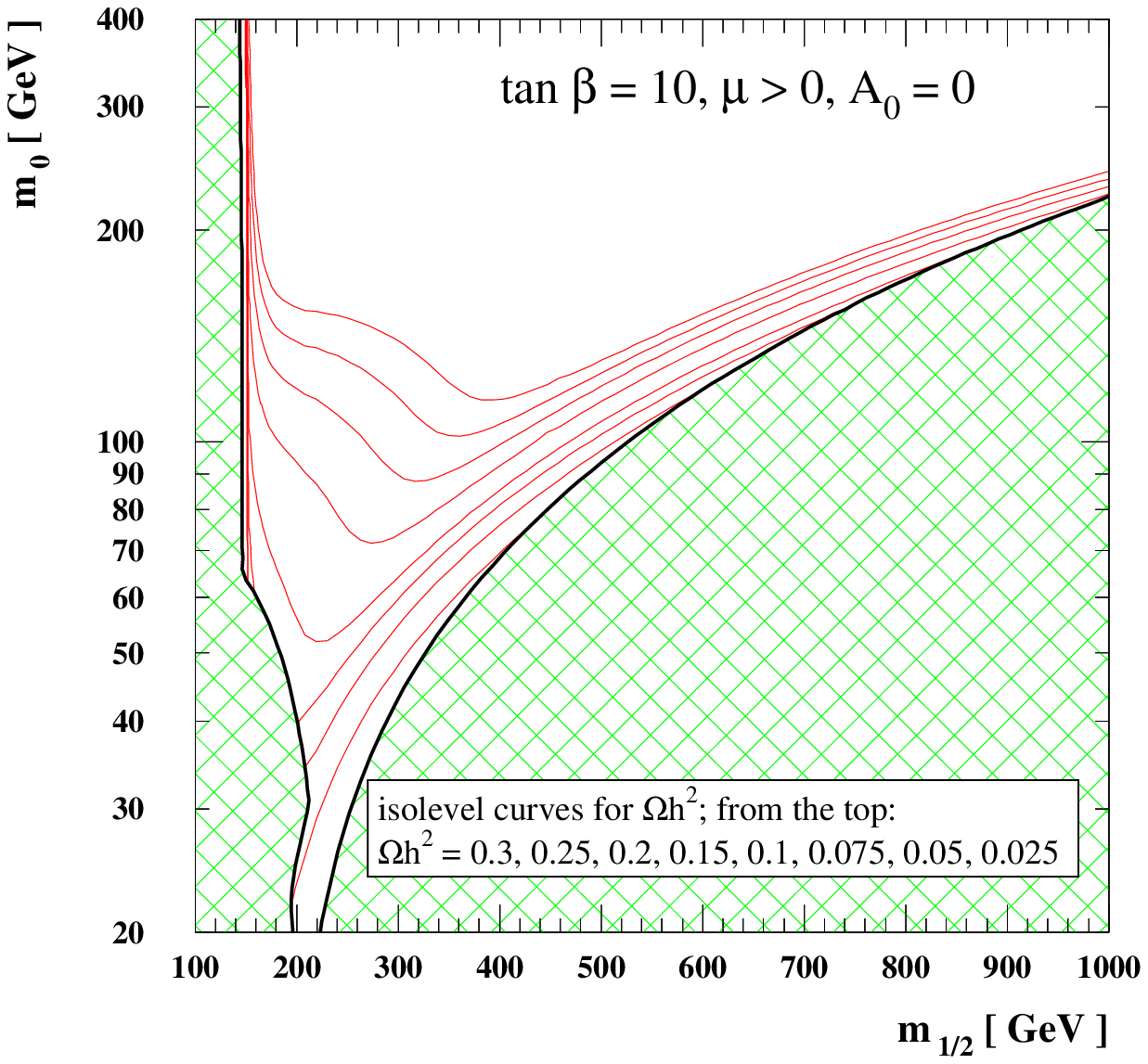,width=0.49\textwidth}}
\centerline{\epsfig{file=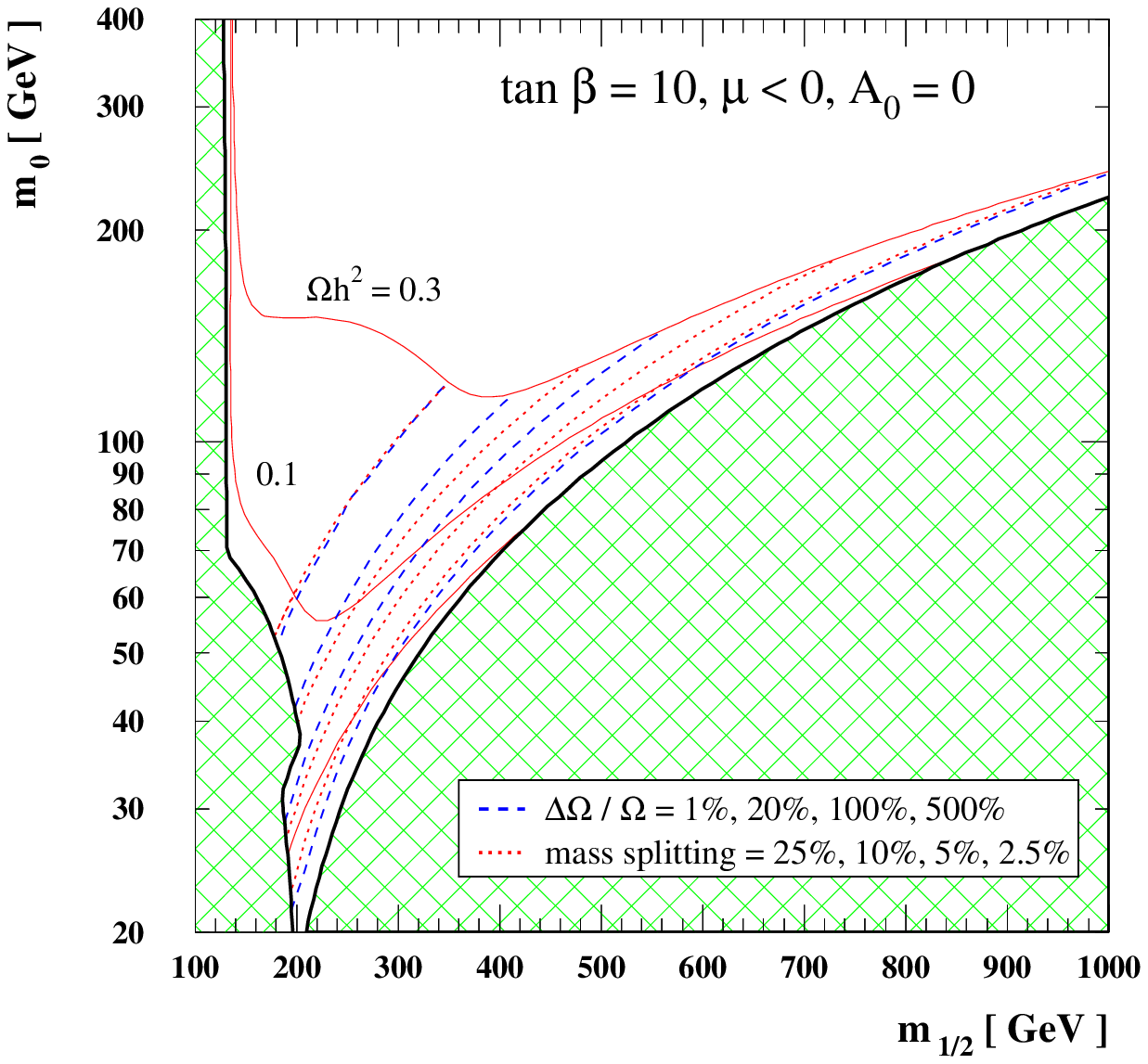,width=0.49\textwidth}
\epsfig{file=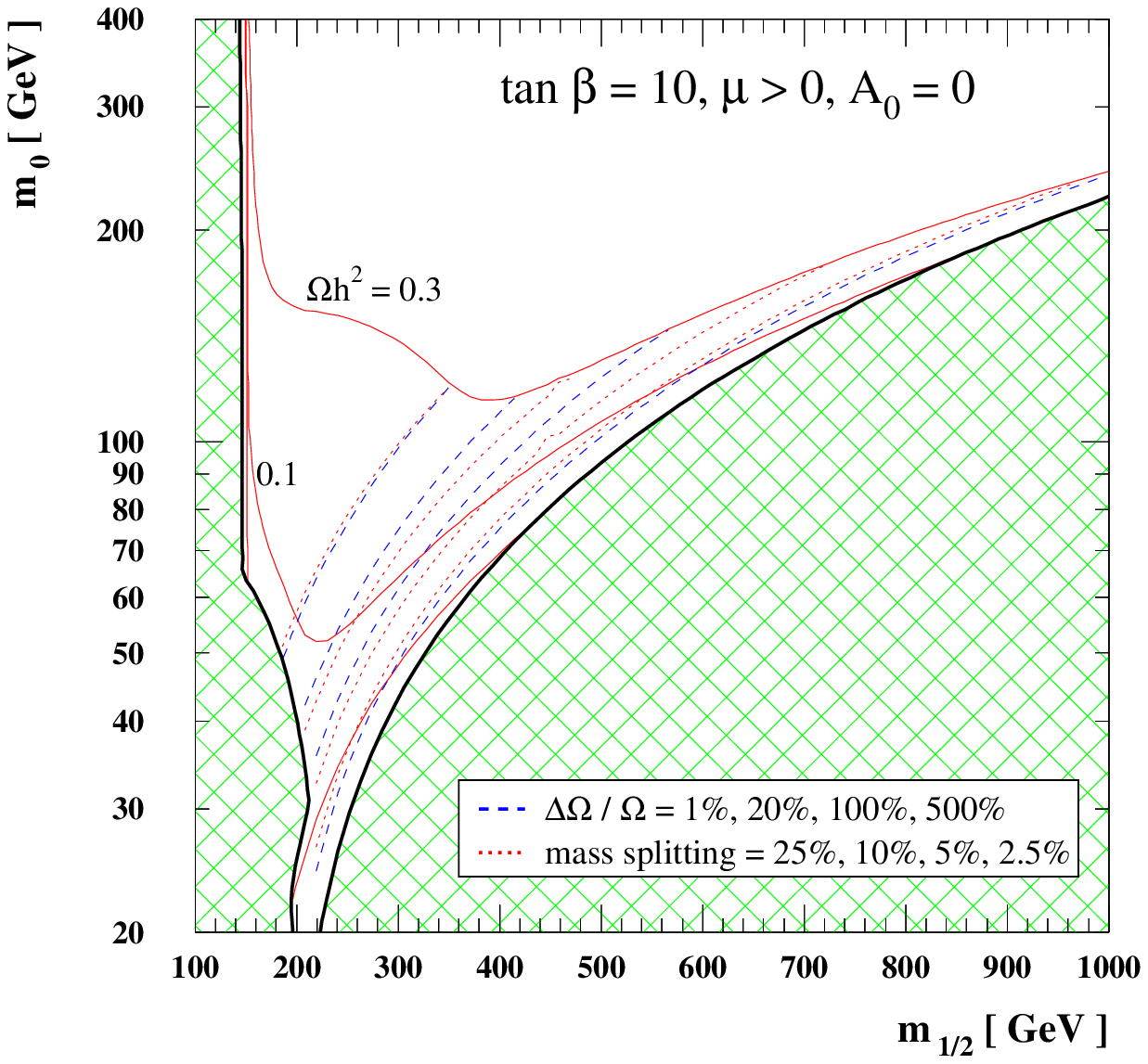,width=0.49\textwidth}}
\caption{Results for $\tan \beta=10$ and $A_0=0$. The isolevel curves for the
relic density $\Omega_\chi h^2$ are shown in the top panels. In the bottom 
panels, curves indicate how big the error on the relic density would be if 
coannihilations were not included. The mass splitting between the lightest 
neutralino and the lightest stau is also indicated.}
\label{fig:stau-tg10}}

We consider first the case $A_0=0$ and $m_0 \lsim m_{1/2}$.  The slepton
coannihilation region was recognised in Ref.~\cite{Ellisstau1} and has been the
focus of several recent studies, including
~\cite{Ellisstau2,GLP,roszkowski,bbb,bbb7.64}.
As an example, we take $\tan \beta=10$ and study the $m_{1/2} - m_0$ plane for
both signs of $\mu$. In the top panels of Fig.~\ref{fig:stau-tg10}, we plot
isolevel curves for the neutralino relic abundance, including coannihilation
effects, for 8 different values of $\Omega_\chi h^2$ starting from $\Omega_\chi
h^2 = 0.3$ and decreasing down to $\Omega_\chi h^2 = 0.025$. The latter value
corresponds to the case in which neutralinos would be a subdominant dark matter
component in the Universe 
but could still account for a major part of the dark matter in
galaxies. The shaded area on the left in each panel is excluded by accelerator
constraints, while the shaded area towards the bottom right corner in each
panel is removed because in this region the LSP is the lightest stau rather
than the lightest neutralino: its upper bound marks the line along which the
(bino-like) neutralino and the lightest stau have equal mass.

We can give a schematic interpretation of the results displayed starting with
the isolevel curves on the top left corner of each panel, where all isolevel
curves converge to a narrow band. There, the model has a relatively heavy sfermion
sector, and the lightest neutralino mass is just a few GeV larger than half the
$Z^0$ boson mass. The bino pair annihilation rate into fermions is dominated by
the diagram with $Z^0$ in the s-channel at energies just slightly displaced
from the $Z^0$ resonance: this resonant annihilation leads to acceptable values
of $\Omega_\chi h^2$ in a narrow band. Thus $\Omega_\chi h^2$ is very sensitive
to the parameter $m_{1/2}$ as the value of the bino mass has to be fine-tuned
in such a way that the thermally averaged cross section picks the right portion
of the resonance; for larger $m_\chi$, $\langle\sigma{v}\rangle$ drops rapidly
and $\Omega_\chi h^2$ becomes large, closer to the resonance
$\langle\sigma{v}\rangle$ becomes exceedingly large and $\Omega_\chi h^2$ very
small. When we follow the isolevel curves from the top left down to smaller
values of $m_0$, sfermion masses decrease and the amplitude for neutralino
annihilations into fermions mediated by sfermion exchange in $t$- and
$u$-channels increase, eventually becoming dominant. The largest contribution
to $\langle \sigma v \rangle$ is given by the $\tau^- \tau^+$ final state which
is mediated by the lightest sfermion, in this case the lightest stau.  Such an
increase in the cross section when moving to lower $m_0$ can be compensated by
increasing the bino mass, i.e. by shifting to larger $m_{1/2}$. This explains
the trend of the isolevel curves starting to align along the diagonal in the figure.
Further down this diagonal the mass of the lightest stau is more or less constant, but the mass of the heaviest stau increases. This increased mass splitting between the staus causes the cross section to increase, but this increase is again compensated by a larger neutralino mass.  Before
reaching the lower right corner where the lightest stau is the LSP, 
coannihilation effects take over:
$\langle\sigma_{\rm{eff}}{v}\rangle$ becomes rapidly dominated by
neutralino-slepton and slepton-slepton contributions. The coannihilation cross
section decreases with increasing initial state masses, an effect which can be
compensated by decreasing the mass splitting between the LSP and the stau. (This
effect is further discussed in connection with Fig.~\ref{fig:slweff}.)
Hence, the isolevel curves bend almost parallel to the bound of the excluded
region, confined to a band which gets progressively narrower towards larger
values of $m_0$ and $m_{1/2}$. As can be seen by comparing the top left with
the top right panel, the flip in $sign(\mu)$ does not alter this overall
picture but just slightly displaces  the position of the isolevel curves in the
$m_{1/2} - m_0$ plane.

The region where coannihilation effects become important is highlighted
in the bottom panels of Fig.~\ref{fig:stau-tg10}. There 
dashed lines show the isolevel curves for the relative difference in relic 
abundance computed neglecting and including coannihilations,
$\Delta \Omega / \Omega \equiv (\Omega_{\chi,~\rm no~coann}-\Omega_{\chi,~\rm coann})/
  \Omega_{\chi,~\rm coann}$.
The values we display span from 1\% to 500\%. Isolevel curves for the 
relative mass splitting between the lightest stau and the lightest neutralino
are shown as well and, as expected, the correlation with the isolevel
curves of $\Delta\Omega/\Omega$ is evident.

In Fig.~\ref{fig:stau-tg30} we repeat the exercise for
$\tan \beta =30$ and the picture we see is analogous to what we found in
the $\tan \beta =10$ case. The shift of the isolevel curves to slightly
larger $m_0$ values is mainly due to the fact that, comparing corresponding
points in the plane, sleptons are driven to slightly lighter masses at larger
 $\tan \beta$.

\FIGURE[t]{
\centerline{\epsfig{file=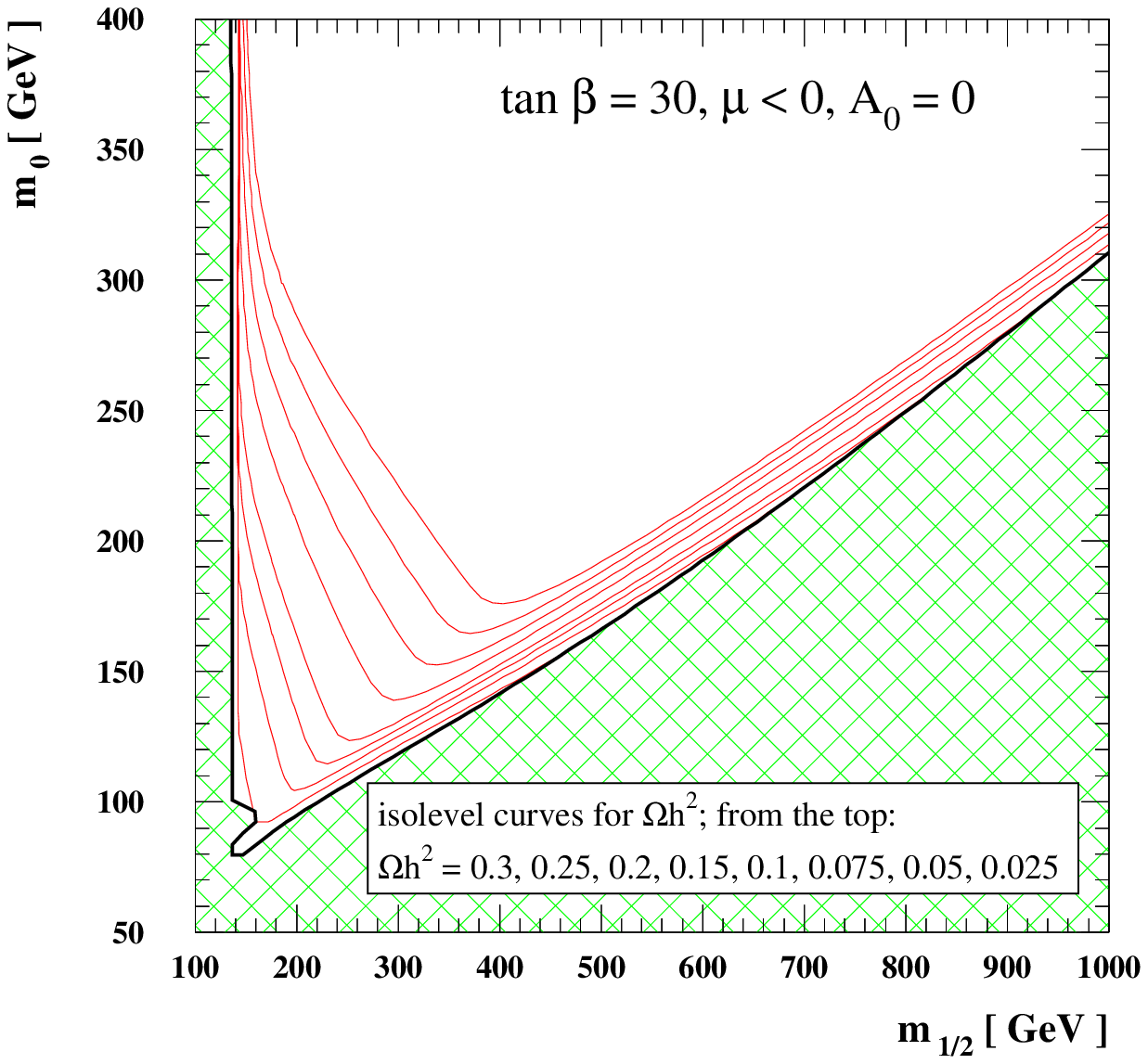,width=0.49\textwidth}
\epsfig{file=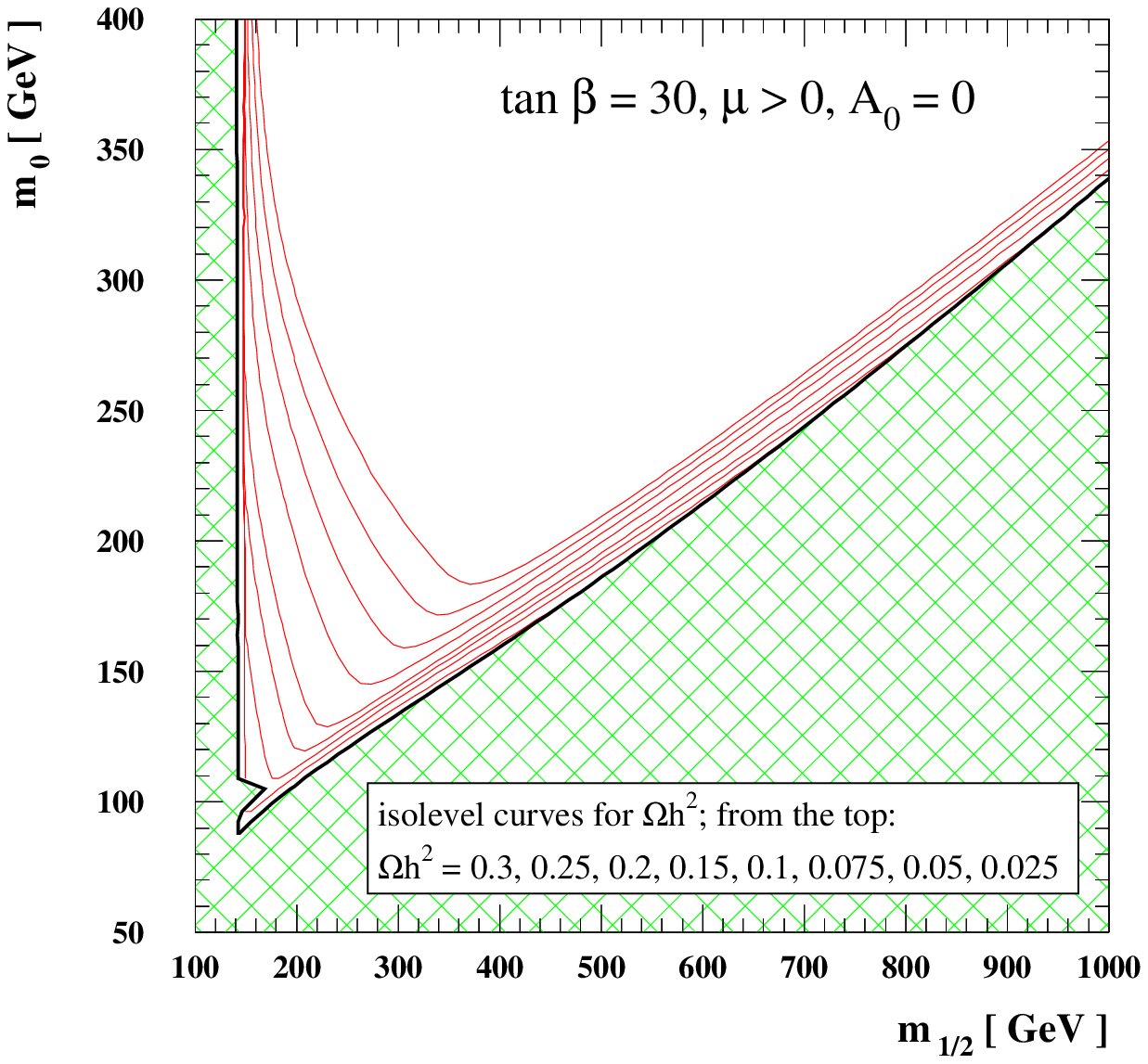,width=0.49\textwidth}}
\centerline{\epsfig{file=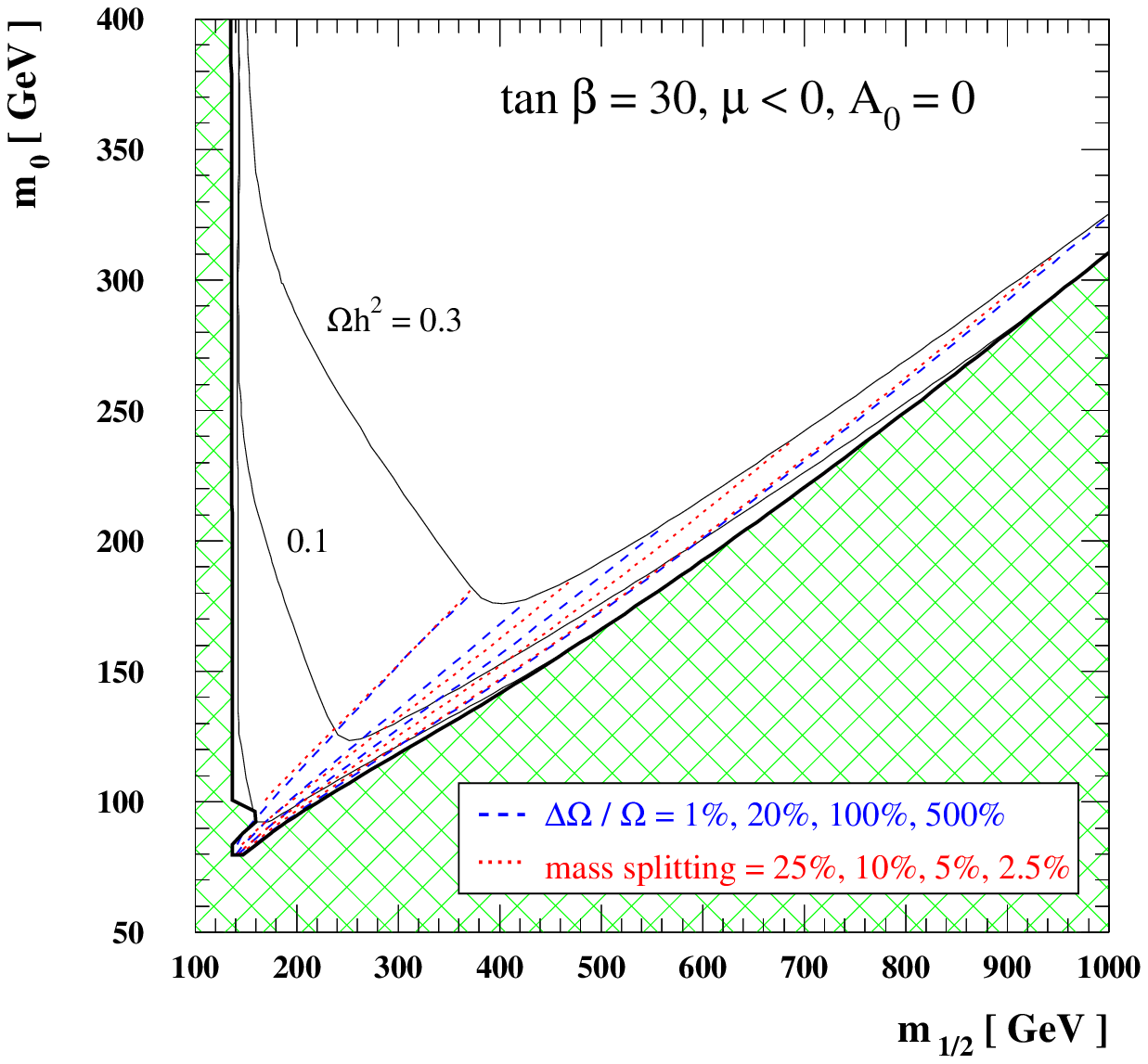,width=0.49\textwidth}
\epsfig{file=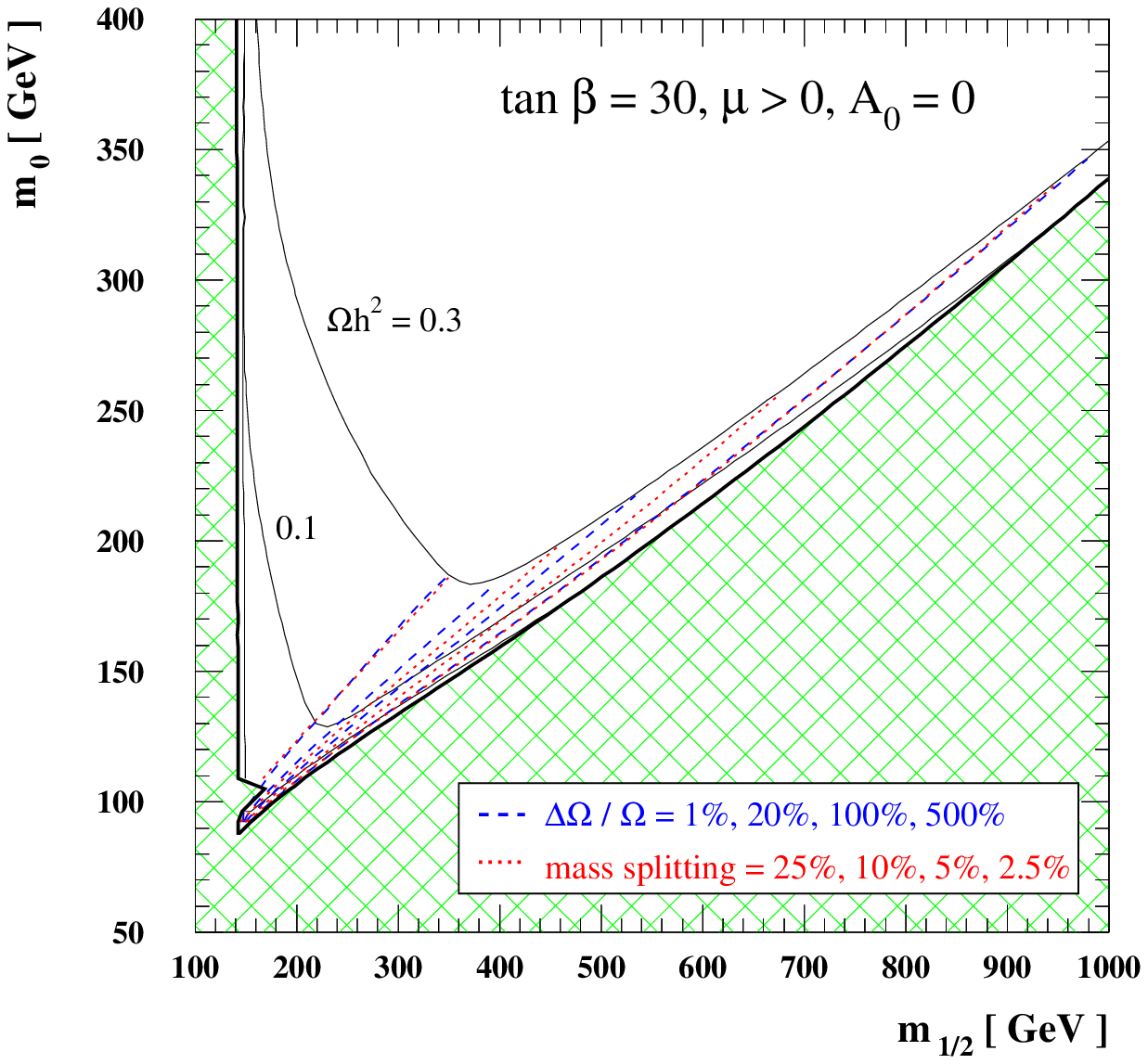,width=0.49\textwidth}}
\caption{Results for $\tan \beta=30$ and $A_0=0$. The isolevel curves for 
the relic density $\Omega_\chi h^2$ are shown in the top panels. In the bottom 
panels, curves indicate how big the error on the relic density would be if 
coannihilations were not included. The mass splitting between the lightest
neutralino and the lightest stau is also indicated.}
\label{fig:stau-tg30}}

From Figs.~\ref{fig:stau-tg10} and \ref{fig:stau-tg30}, as well as from
other sample checks performed for other values of $\tan \beta$,
we can infer that, as a rule of
thumb, in the $A_0=0$ case a 25\% mass splitting between the bino and 
the lightest stau is approximately the borderline below which slepton 
coannihilations have to be included for an estimate of $\Omega_\chi$ with 
a 1\% accuracy level.

\FIGURE[t]{
\centerline{\epsfig{file=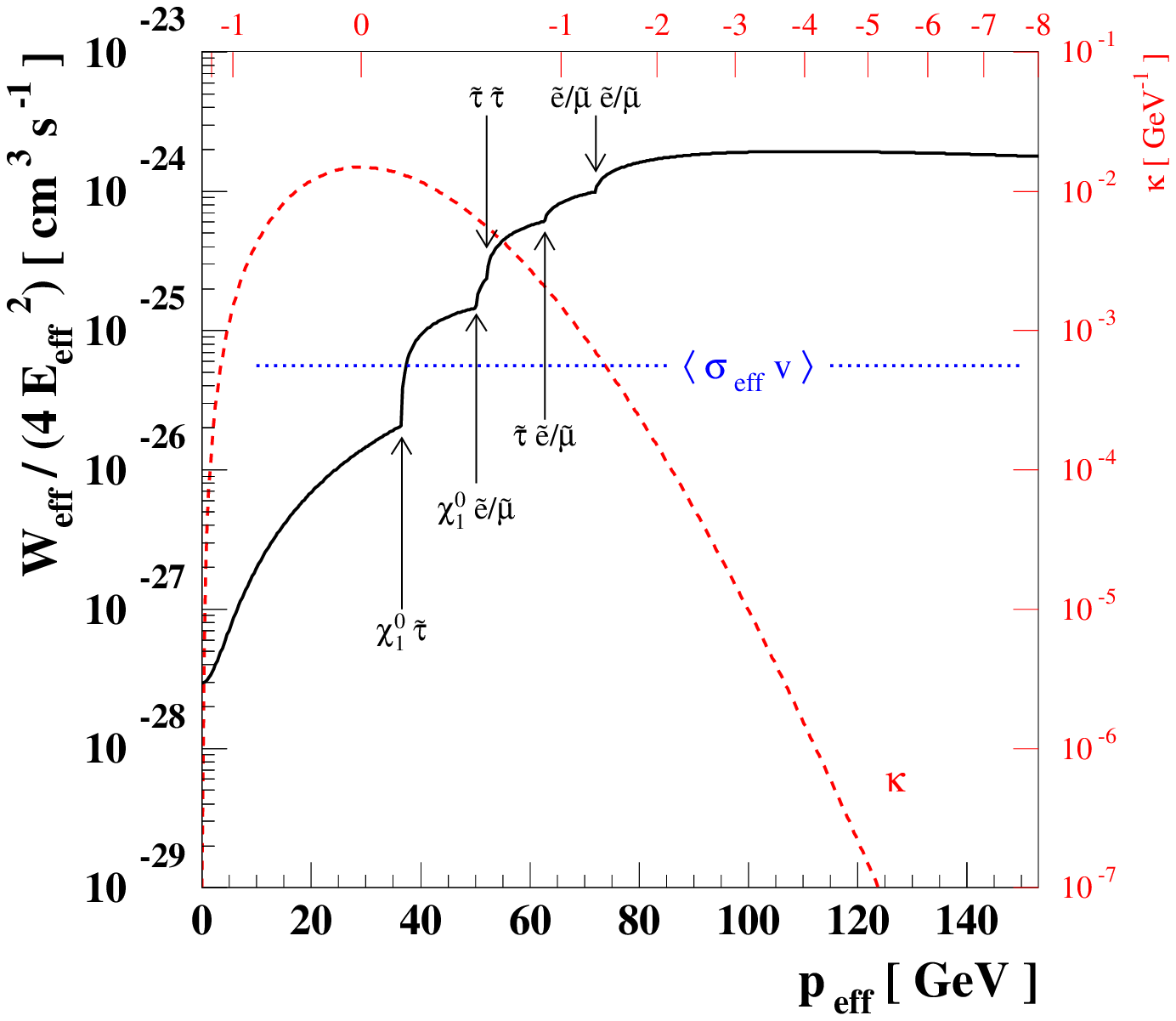,width=0.49\textwidth}
\epsfig{file=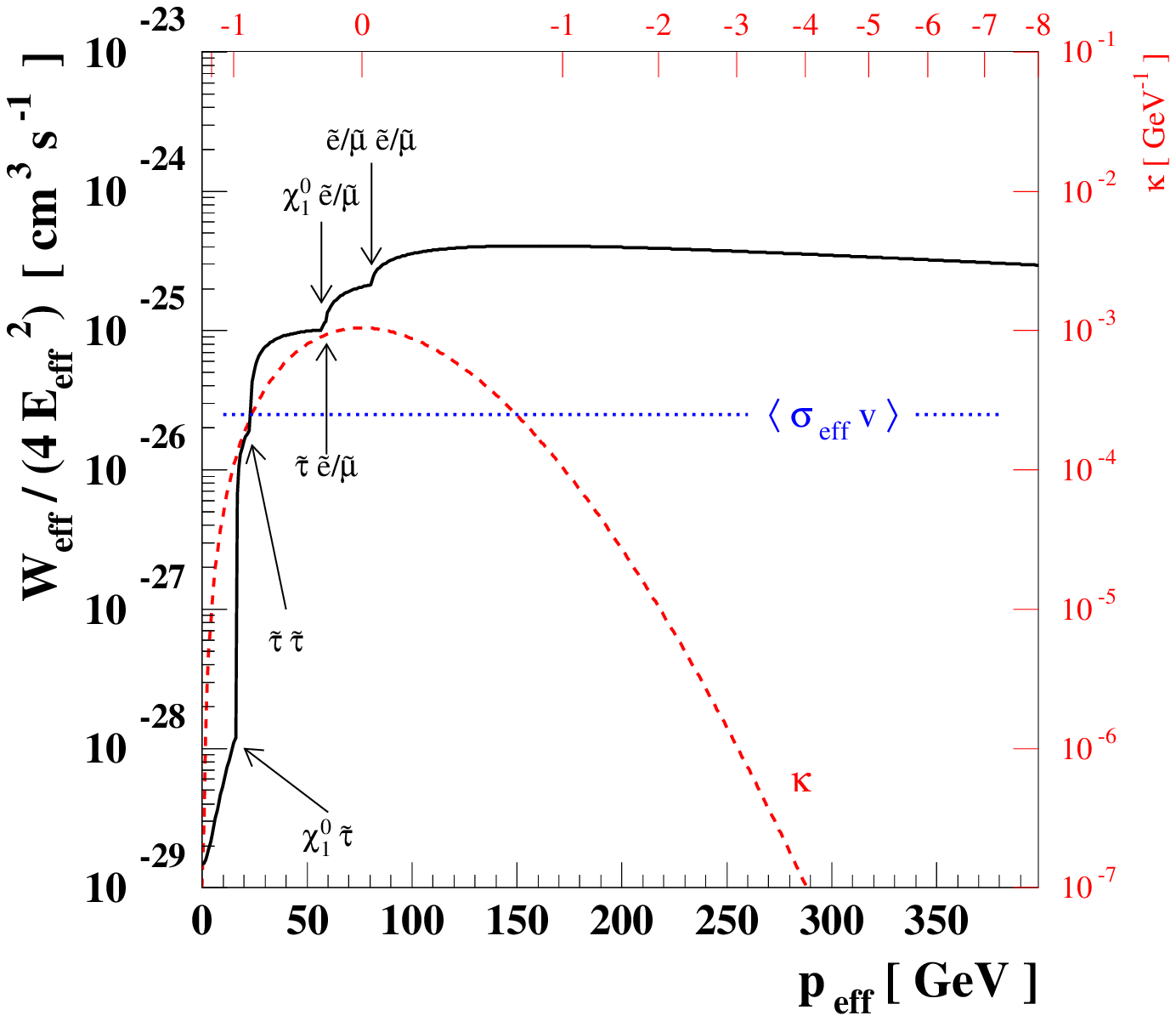,width=0.49\textwidth}}
\caption{The effective annihilation cross section $W_{\rm eff}/4E_{\rm eff}^2$ versus $p_{\rm eff}$ for two sample cases (model E and F in Table~\ref{tab:examples} in Appendix \ref{appexmodels}) with $\tan \beta=10$ and $\Omega_\chi h^2=0.115$.
The figure illustrates how coannihilation effects become more important the further out in the coannhilation tail we get.
See text for a discussion.}
\label{fig:slweff}}

We have briefly mentioned the reason why the stau coannihilation region 
takes the form of a tail extending to large values of $m_{1/2}$. When $m_{1/2}$, 
and therefore $m_\chi$, is increased, it is necessary to increase the effect of 
coannihilations, and therefore to lower the relative mass splitting between the 
LSP and the NLSP, if the relic density should not increase. Fig.~\ref{fig:slweff} 
gives a detailed illustration of this effect. We select here two models with 
$A_0=0$, $\tan \beta =10$, $sign({\mu})$ positive,
and the $m_{1/2}$, $m_0$ pair chosen in such a way that the relic abundance for
both models is $\Omega_\chi h^2 = 0.115$, but $\Delta\Omega/\Omega$ is 100\%
for one model (left panel, model E in Table~\ref{tab:examples}) and 1000\% for
the other (right panel, model F in Table~\ref{tab:examples}). Analogously
to some of the figures in Section~\ref{omegaex}, we plot $W_{\rm{eff}}/4
E^2_{\rm{eff}}$, as well as the weight function $\kappa$ computed at the freeze
out temperature.  Going from the left to the right panel, $m_{\chi}$ increases
from 138.5 GeV to 371.1 GeV, the lightest stau from 148.0 GeV to 371.8 GeV.
The mass splitting consequently goes from 6.8\% to 0.21\%. The increase in
neutralino mass causes a sharp drop in the $\chi^0_1$-$\chi^0_1$ annihilation
cross section\footnote{As $W_{\rm{eff}}/4 E^2_{\rm{eff}}$ at $p_{\rm{eff}}=0$
  is the neutralino annihilation rate at zero temperature, which is the
  relevant quantity for indirect dark matter detection, we find that, roughly
  speaking, it is a factor of 140 harder to detect indirectly the model on the
  right, a factor of 20 due to the reduced value of the cross section, plus a
  factor of 7 due to the square of the ratio of $m_{\chi}$'s (which takes into
  account the neutralino number density scaling for a fixed dark matter mass
  density).}; on the other hand such a drop can be compensated in the thermally
averaged cross section by increasing the role of coannihilating particles.
Even though their annihilation cross sections have also been slightly reduced
by the mass shift, we have forced their contribution to the thermal average to
be larger by shifting them deeply in the region where the weight function
$\kappa$ is large. More precisely, we have required the coannihilation
threshold to move to lower $p_{\rm eff}$, i.e.\ to have a lower mass
splitting.\footnote{ Note that if we had kept the mass splitting fixed and just
  shifted the mass scale, we would have found an equal shift in the position of
  the coannihilation thresholds on the $p_{\rm eff}$ scale. On the other hand,
  the ``width'' in $p_{\rm eff}$ of the weight function $\kappa$ would have
  increased by about the same factor since the freeze-out temperature is
  higher. Hence, the net result would have been that the right-hand panel would
  have looked very similar to the left-hand panel, except for roughly an
  overall stretch along the $p_{\rm eff}$ scale, and for all cross sections
  being lower. We remind the reader that we have fixed the range of the scale
  on the top of each panel (the value on the scale marks the location of the
  momentum $p_{\rm{eff}}^{(n)}$ defined as
  $\kappa(p_{\rm{eff}}^{(n)})/\kappa(p_{\rm{eff}}^{\rm{max}}) = 10^{n}$, where
  $p_{\rm{eff}}^{\rm{max}}$ is the momentum corresponding to the maximum of the
  weight function $\kappa$) and derived the corresponding range in
  $p_{\rm{eff}}$.}
With these changes in the mass splitting, we have found two models
with about the same thermally averaged cross sections, and then, with the 
rule of thumb
$\Omega_\chi h^2 \simeq 10^{-27}$~cm$^3$~s$^{-1}/\langle \sigma_{\rm{eff}}v \rangle$
one can expect similar relic densities. In fact, for the two cases discussed 
here we have chosen the parameters to give exactly the same neutralino 
relic density $\Omega_\chi h^2 = 0.115$ from the full calculation.

\FIGURE[t]{
\centerline{\epsfig{file=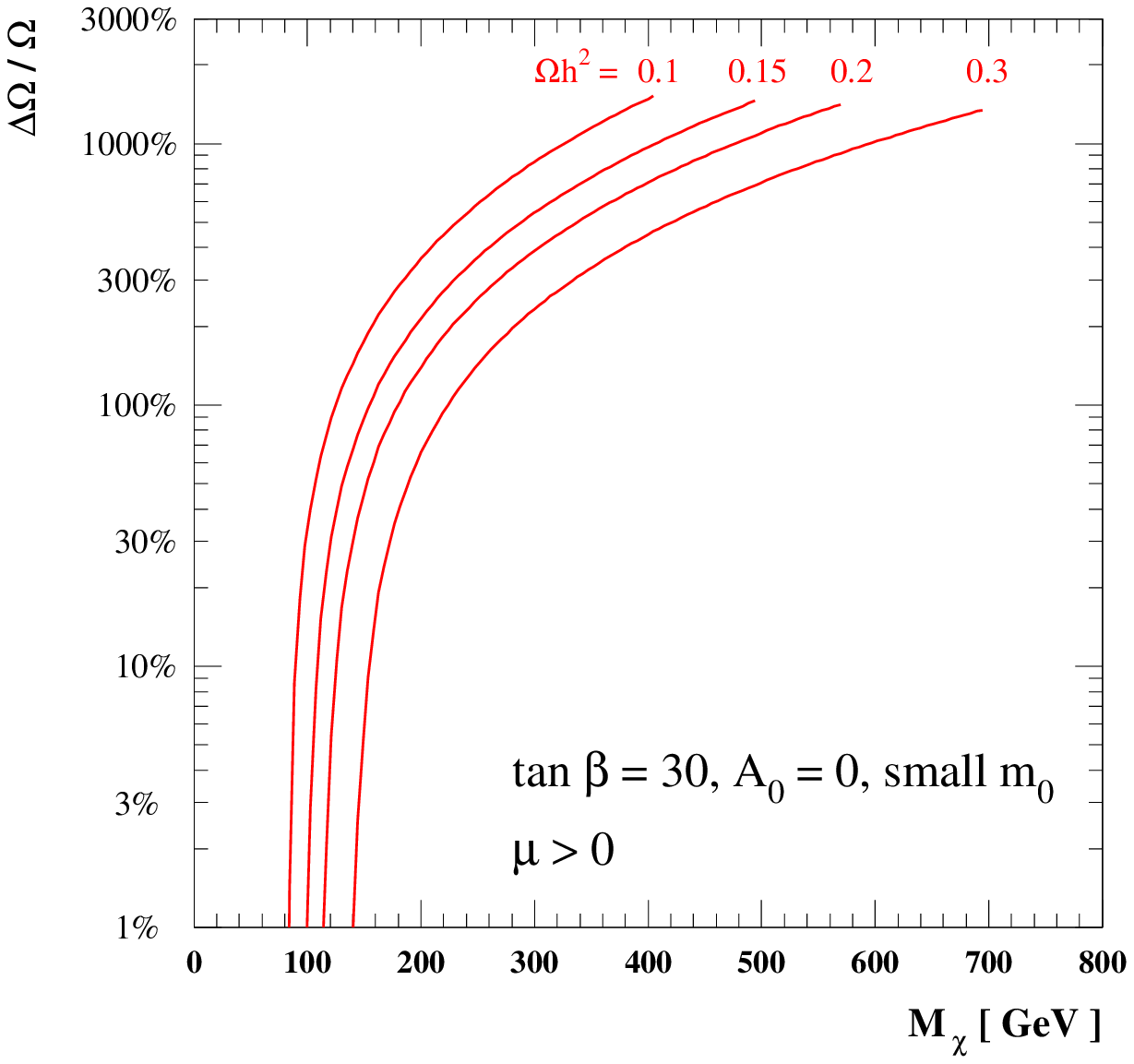,width=0.49\textwidth}
\epsfig{file=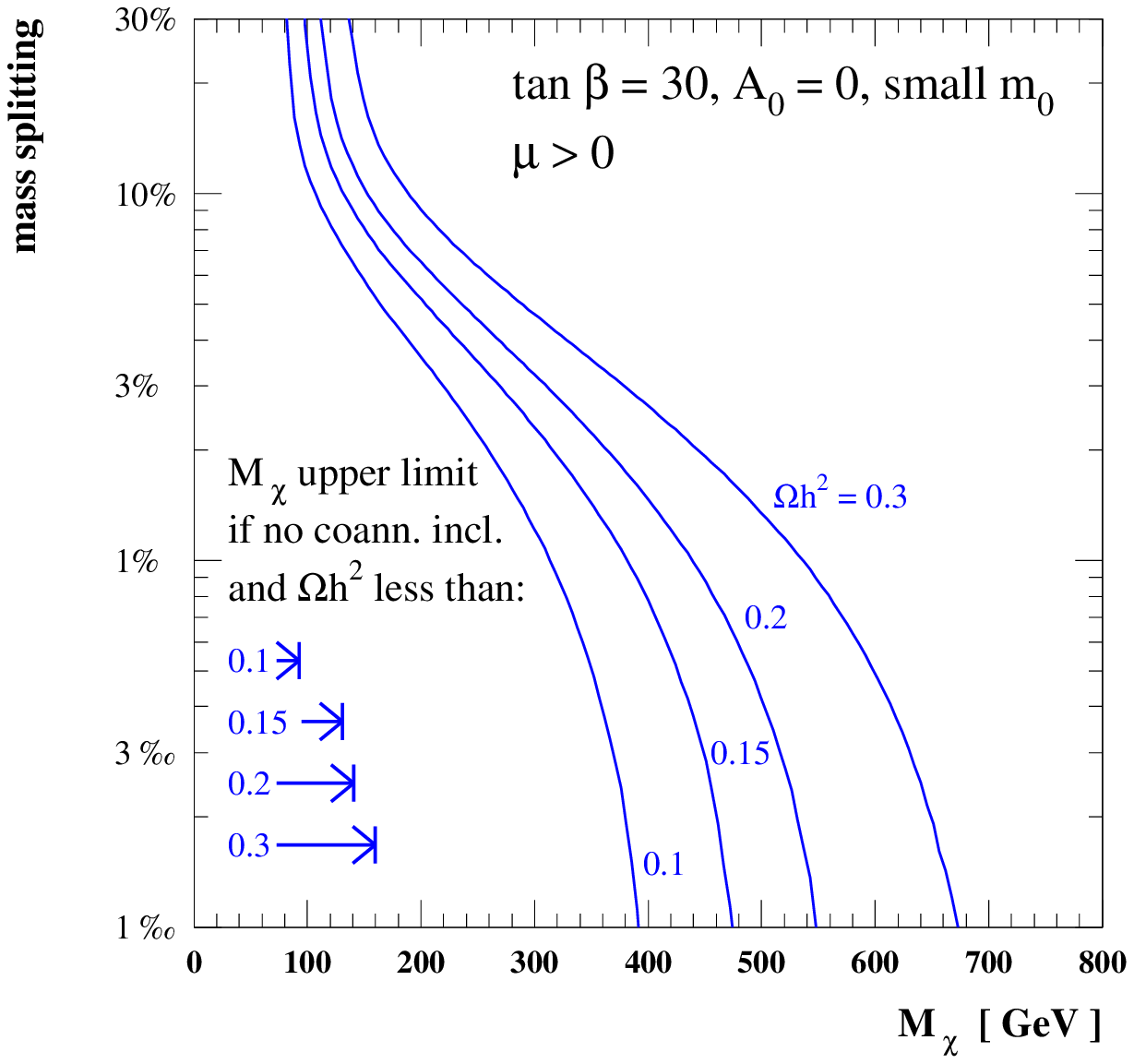,width=0.49\textwidth}}
\caption{For a set of values of $\Omega_\chi h^2$, we show the
  importance, $\Delta\Omega/\Omega$, of including slepton coannihilations, as well as the
  mass splitting between lightest neutralino and lightest stau,
  both as functions of the neutralino mass. We have here chosen 
  $\tan \beta=30$, $\mu >0$, $A_0=0$ and low $m_0$ 
  (the latter to be in the stau coannihilation region). In the right-hand panel,
  we also indicate by arrows where the upper limit on the mass would be if
  coannihilations were not included (for these limits the vertical axis is meaningless). 
  The shift to higher masses when coannihilations
  are included is clearly seen.}
\label{fig:stau-mdiff}}

From the above discussion it should also be evident that we cannot play the
same game to arbitrarily large neutralino masses. At some point one reaches the
edge at which the role of coannihilating particles cannot be further
strengthened and one finds an upper bound on the neutralino mass (for that
specific choice of $\tan\beta$, $sign({\mu})$ and $\Omega_\chi h^2$). The 
introduction of slepton coannihilations have extended the cosmologically 
allowed mass interval of the bino-like LSP in the mSUGRA context. This fact 
has already been pointed out by several authors. In the work of some authors 
\cite{bbb,bbb7.64} it seems that the neutralino mass might be unbounded from above, 
but the majority \cite{roszkowski,Ellisstau1,Ellisstau2} found that there was 
still an upper limit to the allowed mass. Given the high precision 
in the relic abundance calculation we present in this analysis,
we are able to make a qualitative analysis of the high $m_\chi$ region, and indeed 
we find upper limits to the neutralino mass. We will show our results in the
case $\tan \beta = 30$ and positive $\mu$ considered in Fig.~\ref{fig:stau-tg30}. 
Keeping in mind that we are in the $m_0 \lsim m_{1/2}$ regime, there is a one to 
one correspondence between the pair ($m_{1/2}$,$m_0$) and the pair 
($m_\chi,(m_{\tilde{\tau}}- m_{\chi})/m_{\chi}$) or 
($m_\chi$,$\Delta\Omega/\Omega$). We therefore show first in 
Fig.~\ref{fig:stau-mdiff}a the difference in relic density, 
$(\Omega_{\chi,\, \rm no\, coann}-\Omega_{\chi,\, \rm coann})/
\Omega_{\chi,\, \rm coann}$, versus the neutralino mass for a few 
values of $\Omega_\chi h^2$. We clearly see the importance of stau 
coannihilations in the high mass region. In Fig.~\ref{fig:stau-mdiff}b we 
instead show the relative neutralino-stau mass splitting 
(i.e. $\Delta m = (m_{\tilde{\tau}}- m_{\chi})/m_{\chi}$) as a function 
of $m_{\chi}$. Also shown in the figure 
are the upper limits on $m_{\chi}$ for the case where $\Omega_\chi h^2$ is computed
ignoring coannihilation effects; the shift to much larger values, when 
coannihilation effects are included, is evident, as well as the fact
that we do find a new
maximum value of $m_{\chi}$. The results for negative $\mu$
are very similar, while those for $\tan \beta = 10$ are analogous, but show
slightly more stringent upper bounds on the neutralino mass.

\subsection{Chargino coannihilations}

In the focus point region, the value of the soft mass parameter (at the electro-weak scale) for the Higgs doublet that couples to up-type quarks, $m_{H_u}$, is naturally of the electro-weak scale, regardless of $m_0$ \cite{focus-point}. As a consequence, the parameter $\mu$ is forced to be light, and can be at the level of the gaugino mass parameter $m_{1/2}$ or even lower. This implies that the
neutralino LSP may have a large Higgsino fraction and be nearly degenerate in
mass with the lightest chargino 
and the next-to-lightest neutralino. Especially at higher $m_0$-values, the Higgsino fraction can be very large, close to one. Hence, in this high $m_0$ focus point region, chargino (and neutralino) coannihilations are expected to be important.
Chargino
coannihilations have been extensively studied in the generic MSSM
context~\cite{MY,DN,eg-coann}, but have been rarely stressed in the mSUGRA
framework (although they are included in some recent analyses, 
e.g.~\cite{bbb,bbb7.64,bhn-gaugino}).

\FIGURE[t]{
\centerline{\epsfig{file=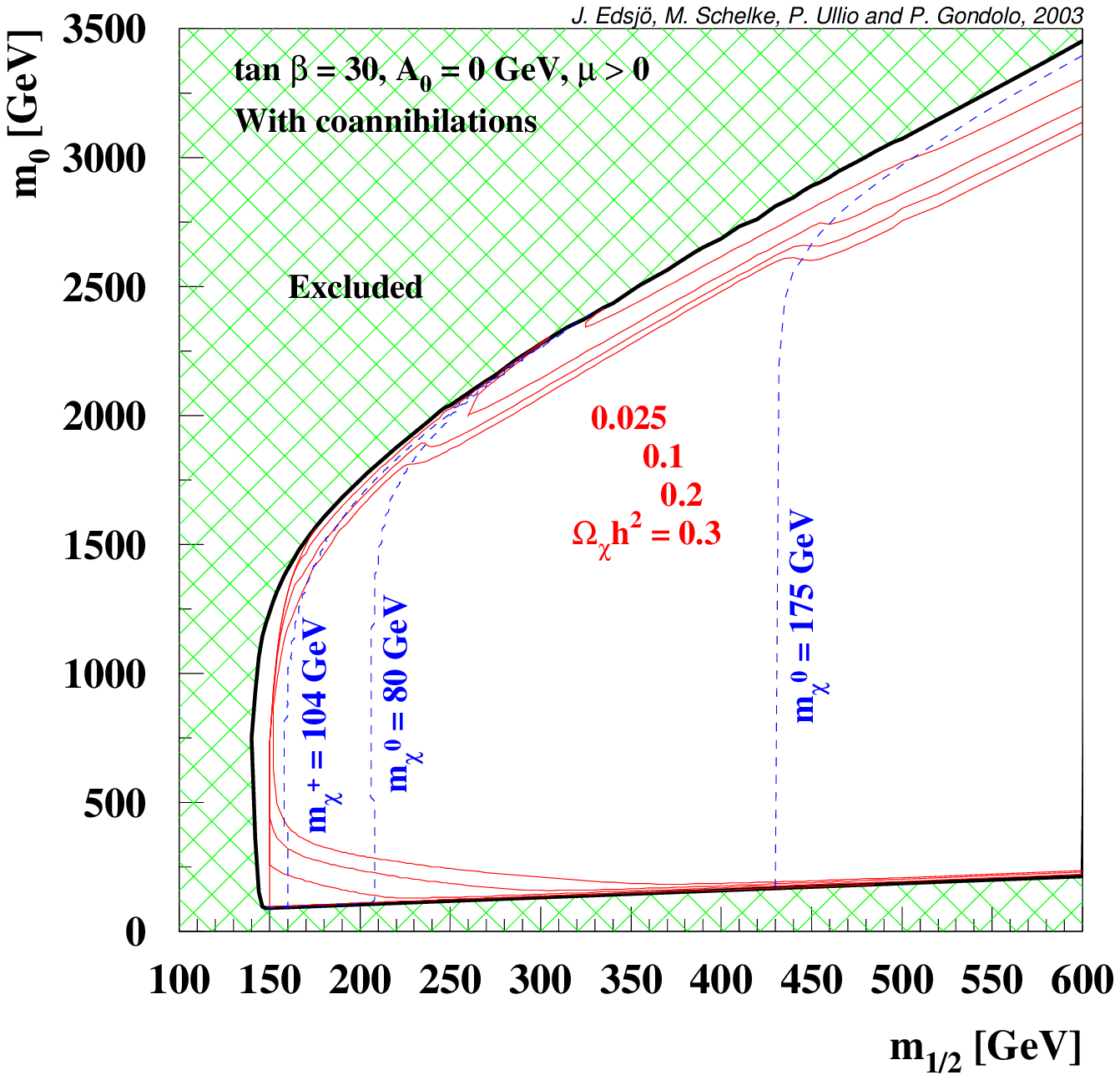,width=0.49\textwidth}
\epsfig{file=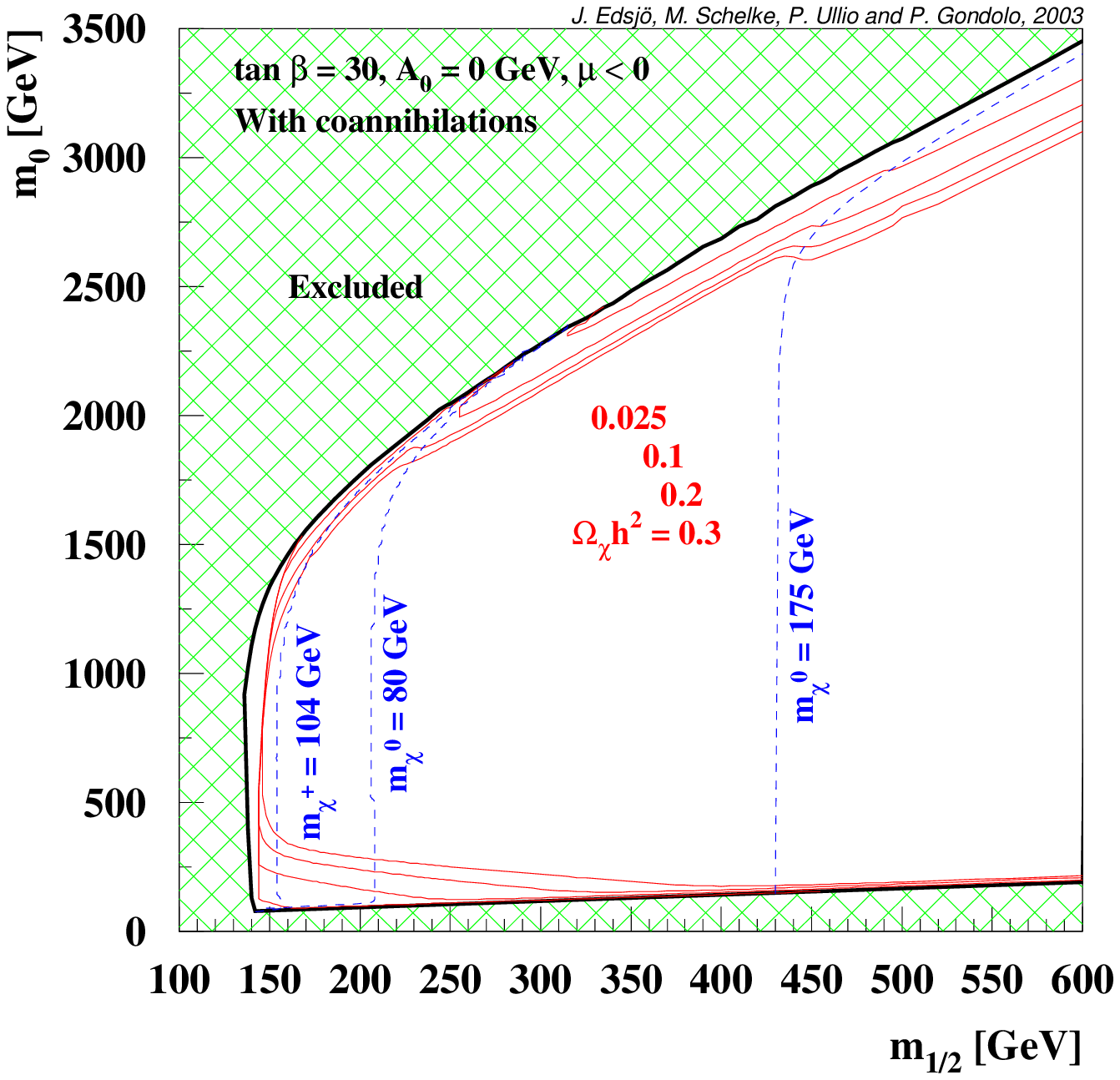,width=0.49\textwidth}}
\caption{The relic density contours (solid lines) for models in the focus point region; $\tan \beta=30$ and $A_0=0$. In a) $\mu>0$ and in b) $\mu<0$.
The kinematic chargino mass limit of 104 GeV and the $W^+W^-$ and $t\bar{t}$ thresholds are indicated.}
\label{fig:chargino-tg30}}

\FIGURE[t]{
\centerline{\epsfig{file=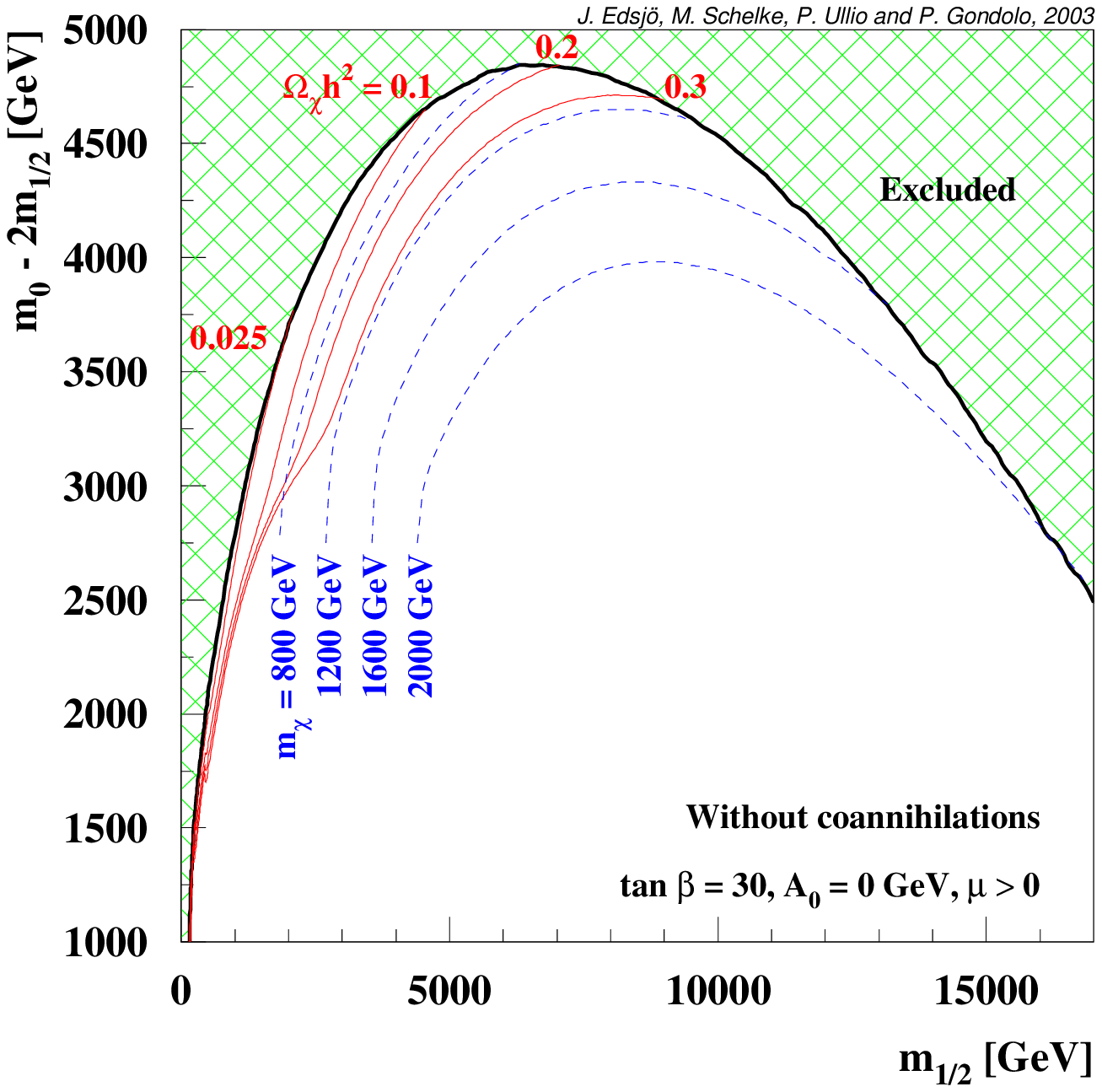,width=0.49\textwidth}
\epsfig{file=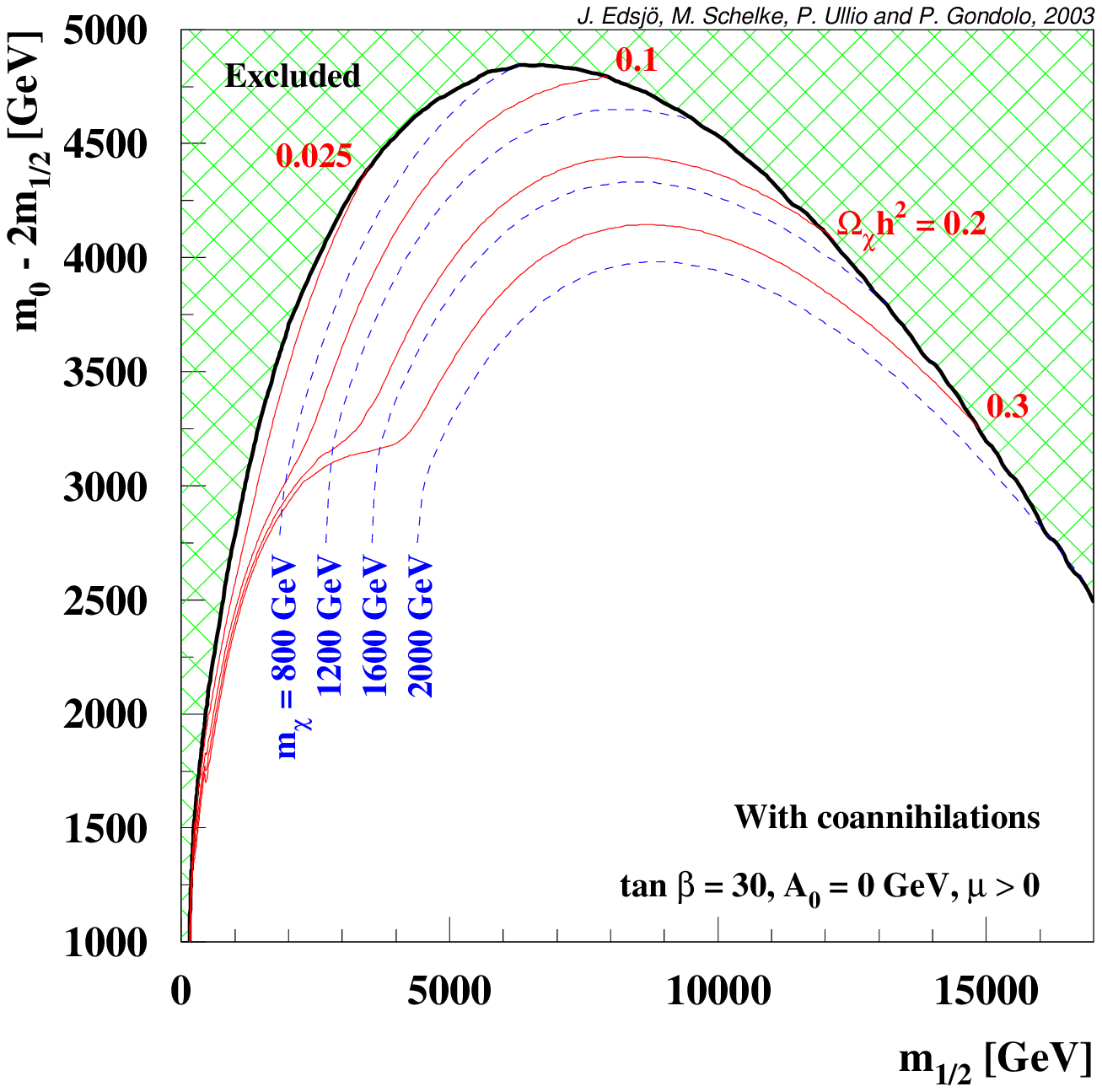,width=0.49\textwidth}}
\caption{The relic density contours (solid lines) for high mass models where chargino coannihilations 
are important; $\tan \beta=30$, $A_0=0$ and $\mu>0$. In a) coannihilations are not 
included, whereas they are included in b). Neutralino mass contours are shown with dashed lines.}
\label{fig:chargino-tg30-high}}

\FIGURE[t]{
\centerline{\epsfig{file=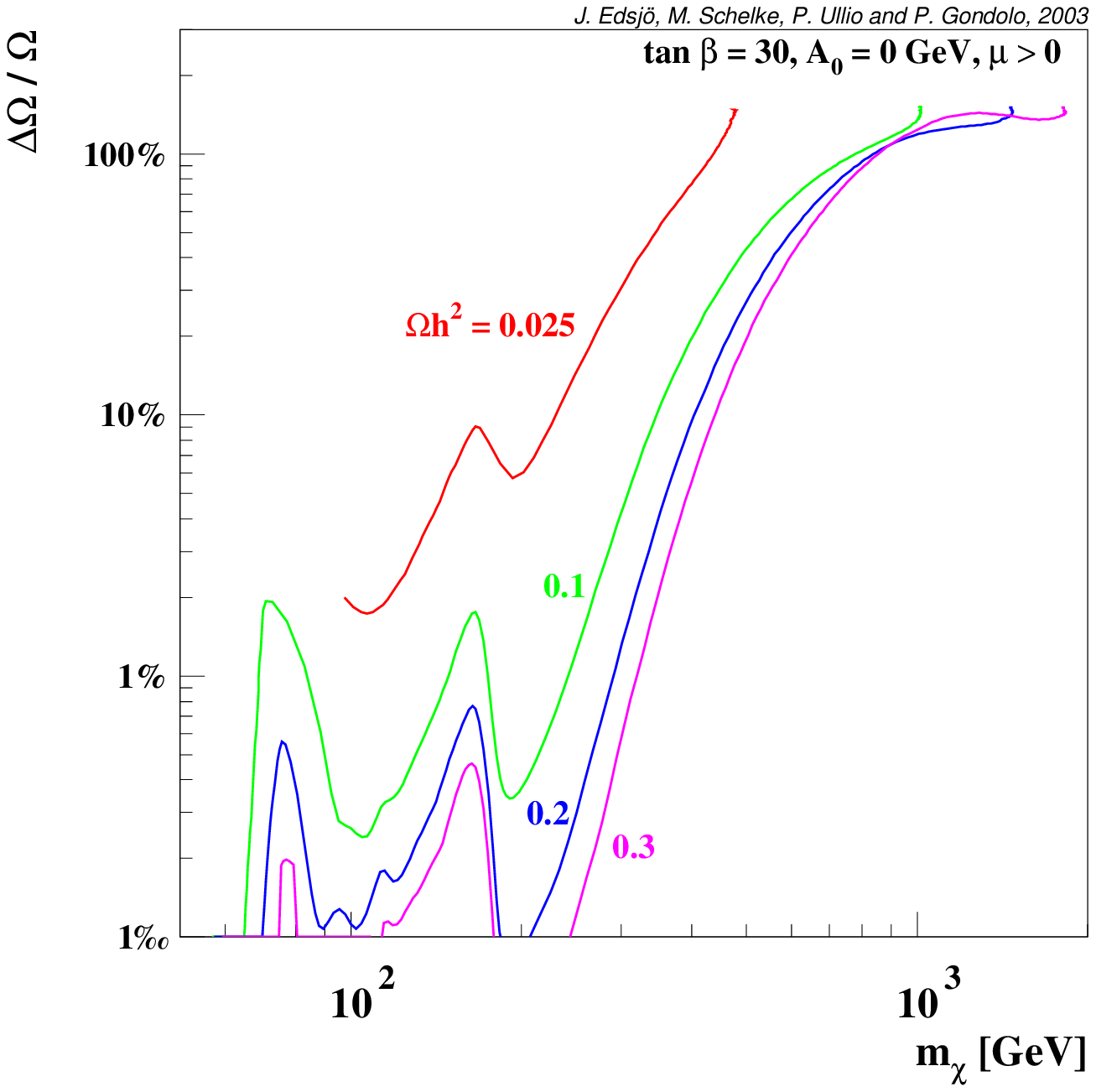,width=0.49\textwidth}
\epsfig{file=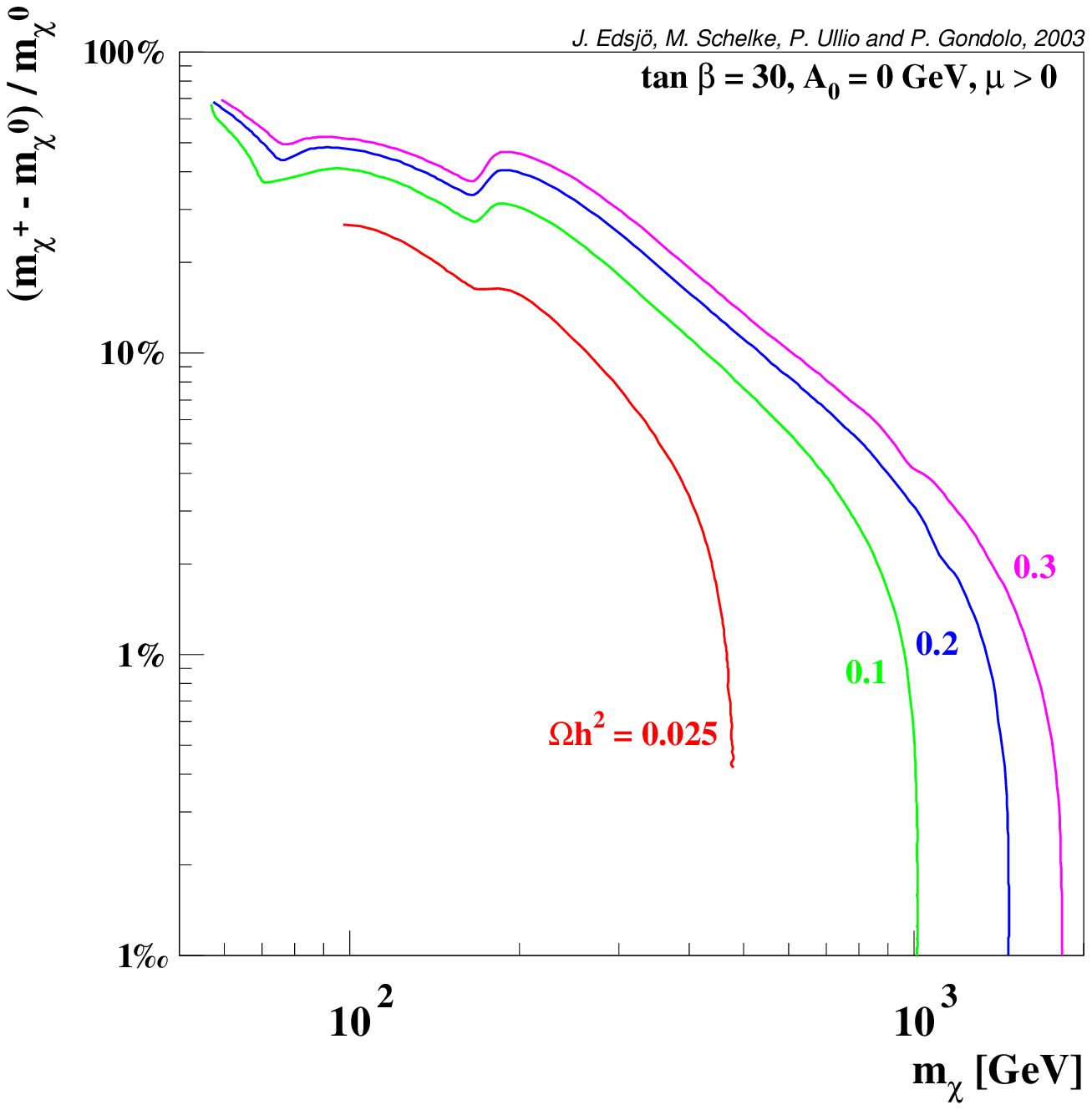,width=0.49\textwidth}}
\caption{We here show the effect of coannihilations for models where chargino coannihilations are important; $\tan \beta=30$, $\mu>0$ and $A_0=0$.
We show in a) $\Delta \Omega / \Omega \equiv (\Omega_{\rm no~coann}-\Omega_{\rm coann})/
  \Omega_{\rm coann}$ versus 
the neutralino mass and in b) the mass splitting between the lightest chargino and 
the lightest neutralino versus the neutralino mass.}
\label{fig:chargino-mupos}}

In Fig.~\ref{fig:chargino-tg30} we show the lower part of the focus point region
for $\tan
\beta=30$ and $A_0=0$. The top-left corners of these figures are excluded due
to no radiative electro-weak symmetry breaking, but close to that region, we
see the focus point region emerge. 
In this region, the Higgsino fraction is usually small, but non-negligible, and the same is true for the effect of chargino coannihilations. This is the part of the focus point region most often discussed in the literature. However, if we continue to higher masses, we get a band of cosmologically interesting relic densities where the Higgsino fraction increases as we go up in mass (at the highest masses, it is close to 1). In this case, coannihilations with the lightest chargino (and the next-to-lightest neutralino(s)) occur and are important.
In Fig.~\ref{fig:chargino-tg30-high}a we show the relic density isolevel curves 
without including coannihilations and in b) with coannihilations included in this high-mass region. 
(Note the scale $m_0-2m_{1/2}$ on the $y$-axis, which is chosen to clearly show the coannihilation region.) We 
clearly see the effect of coannihilations pushing the cosmologically allowed 
region to higher masses. Note that in the focus point region of the parameter space, 
there is no longer a simple relation between the neutralino mass and $m_{1/2}$. 
Some neutralino mass contours are therefore indicated in the figures.

As shown in Fig.~\ref{fig:chargino-mupos}, we find that (just as for the MSSM case \cite{eg-coann}) chargino coannihilations are non-negligible below the $W$ mass where the dominant neutralino-neutralino
annihilation channel is annihilation into fermion anti-fermion pairs via
$s$-channel $Z$-boson exchange. The coupling for a Higgsino-like neutralino 
to the $Z$-boson is very suppressed though, whereas that of charginos is not 
and coannihilations could then give a big effect. However, in the mSUGRA framework, the neutralinos below the $W$ mass are rather mixed than Higgsino-like and the effect is not as dramatic as in the MSSM.
Above the $W$ mass, where 
annihilation into $W^+ W^-$ dominates  for neutralino-neutralino annihilation 
(annihilation into $Z^0 Z^0$ is also significant above the $Z$ mass), 
the effect of coannihilations is smaller
since the annihilation rate into $W^+W^-$ is comparable for neutralinos and
charginos. Still, annihilation into fermion anti-fermion pairs is not
suppressed for chargino-chargino annihilation and thus the chargino-chargino
annihilation cross section is typically slightly higher than the
neutralino-neutralino one. Hence, coannihilations do change the relic density
even for higher masses, and the further up in mass (and down in mass splitting)
we go, the more important they are.
In Fig.~\ref{fig:chargino-mupos}a we plot $\Delta\Omega/\Omega$ versus the
neutralino mass. We see that $\Delta\Omega/\Omega$ is of
the order of 1\% for models with $m_\chi$ just below the $W$ mass and 
drops when we get above $m_W$. Then, $\Delta\Omega/\Omega$ slowly increases 
as we go up in mass with a small drop at $m_\chi=175$ GeV, where the neutralino 
annihilation into $t\bar{t}$ becomes significant, thereby reducing the importance of
coannihilations somewhat. In Fig.~\ref{fig:chargino-mupos}b we show the
relative mass splitting between the lightest chargino and the lightest neutralino 
versus the neutralino mass. We here clearly see the same 
effect as we have seen with slepton coannihilations, i.e.\ as we go up in mass, 
we need to make the mass splitting smaller to maintain the same value of the relic 
density. As for slepton coannihilations, we cannot continue this game to 
arbitrarily high masses and as we see in Fig.~\ref{fig:chargino-mupos}b, we get upper 
limits on the neutralino mass for a given relic density. The corresponding curves for 
$\mu<0$ are very similar and we do not show them separately.

\subsection{Stop coannihilations}

\FIGURE[t]{
\centerline{\epsfig{file=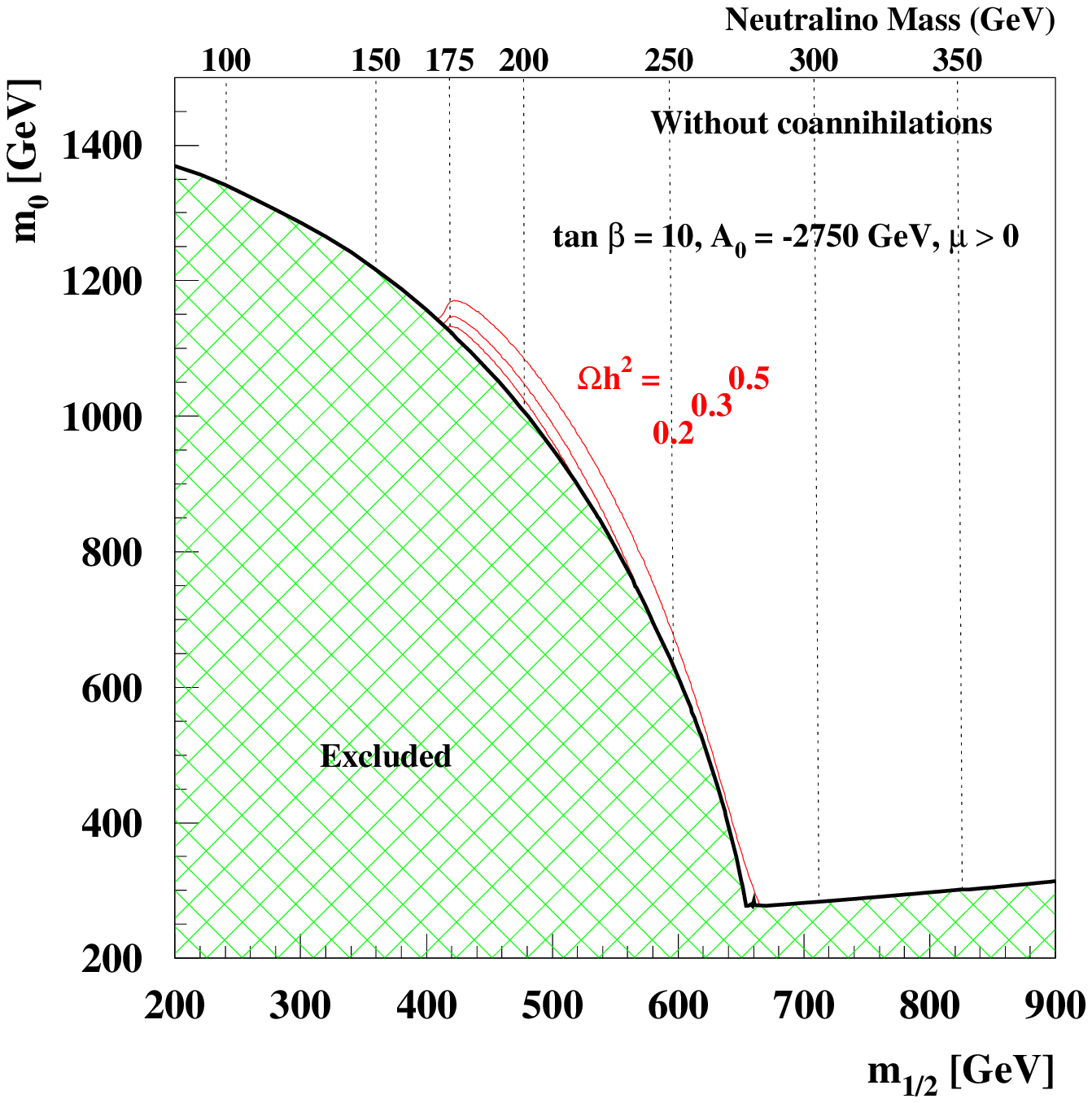,width=0.49\textwidth}
\epsfig{file=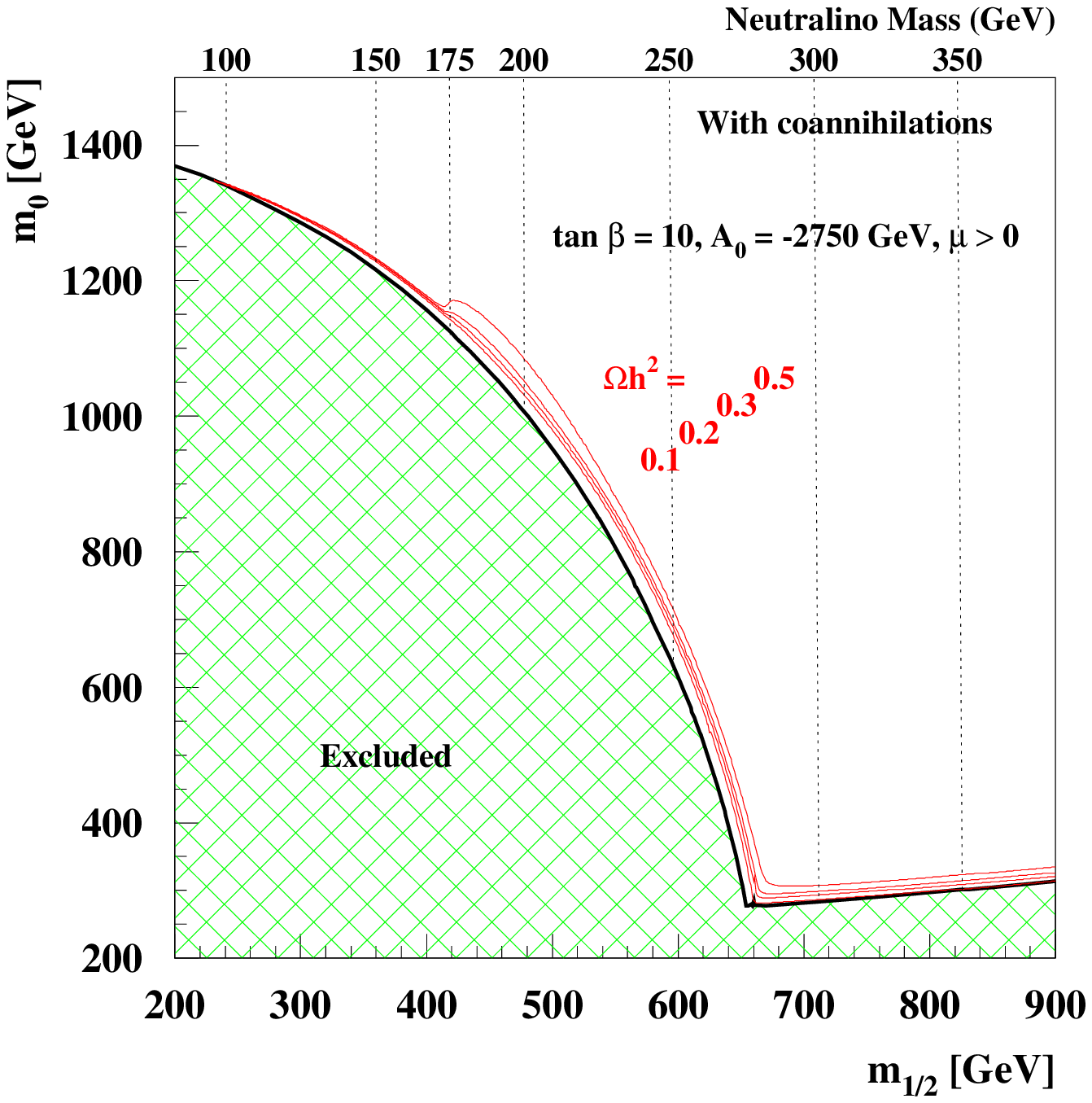,width=0.49\textwidth}}
\caption{Isolevel curves for the relic density a) without
  coannihilations and b) with coannihilations for an example of mSUGRA
  parameters where stop coannihilations are important. In the hatched
  region to the left, $\tilde{t}_1$ is the LSP (except for a narrow 
  band at the lower part of the right edge where the neutralino 
  is the LSP, but where 
  the $H_2^0$ mass constraint is not fulfilled). In the hatched region 
  at the bottom right, $\tilde{\tau}_1$ is the LSP. In the almost
  vertical band of interesting relic densities, a light $\tilde{t}_1$
  is important in two respects, both by boosting the
  $\chi_1^0-\chi_1^0$ annihilation above the $t\bar{t}$ threshold 
  (as seen in a)) and by coannihilations (as seen in b)). 
  The most prominent effect of the $\tilde{t}_1$
  coannihilations is the narrow band for neutralino masses less
  than $m(t) \sim 174$ GeV.}
\label{fig:stop-tg10}}

\FIGURE[t]{
\centerline{\epsfig{file=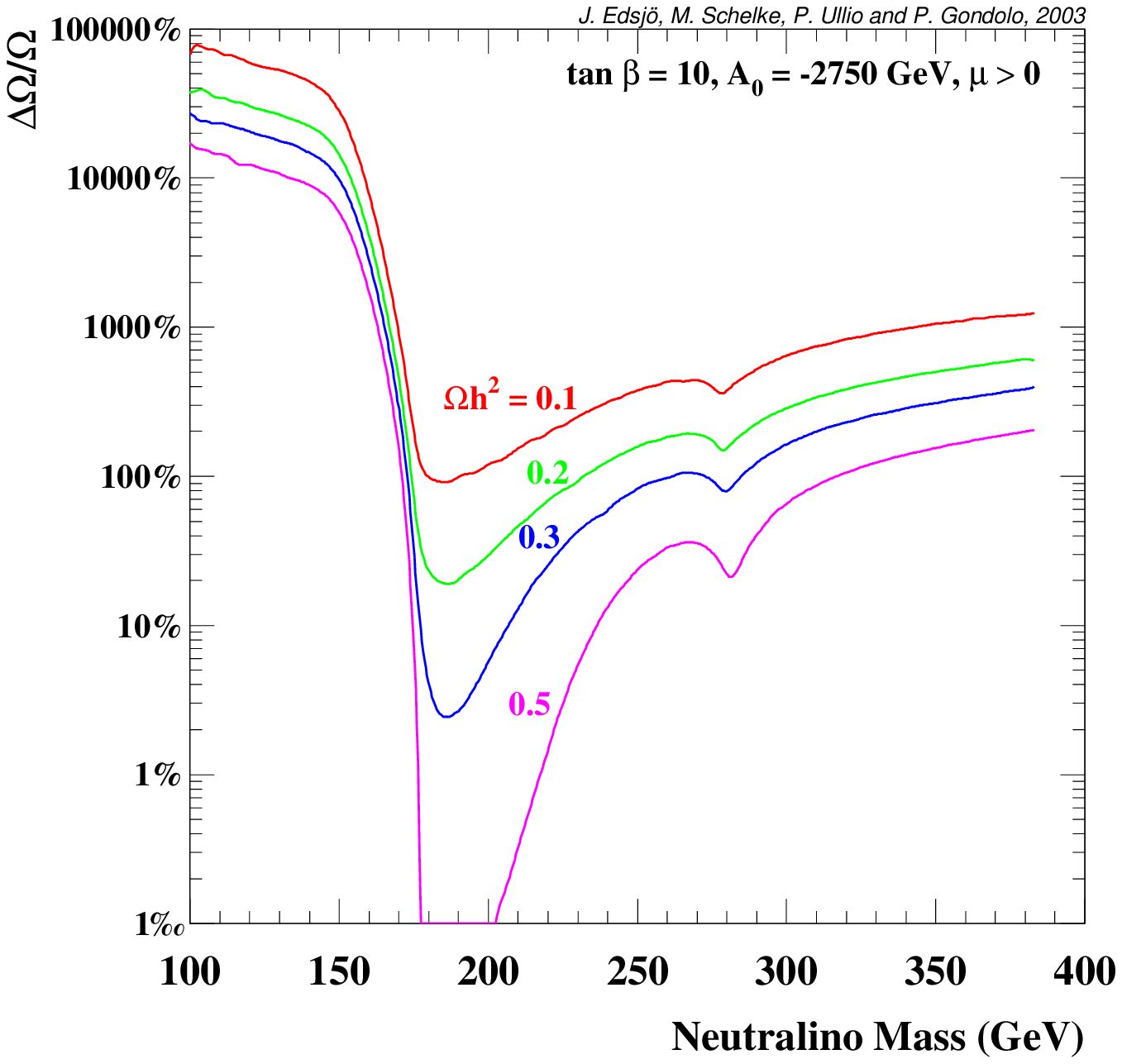,width=0.70\textwidth}}
\caption{For the same set of models as in Fig.~\ref{fig:stop-tg10}, we
  plot $\Delta \Omega / \Omega \equiv (\Omega_{\rm no~coann}-\Omega_{\rm coann})/
  \Omega_{\rm coann}$ versus the neutralino mass. We clearly see the effect that below
  $m_t$, we only get interesting relic densities if coannihilations
  with $\tilde{t}$ are included. When we get close to $m_t$, though,
  even $\chi_1^0 \chi_1^0$-annihilation gets efficient and the
  relative effect of including coannihilations go down. When
  coannihilations with $\tilde{\tau}$ get important at higher masses,
  the relative difference in $\Omega_\chi$ goes up again. The small dip at
  $m_\chi \simeq 280$ GeV corresponds to the low-$m_0$ corner where
  the curves in Fig.~\ref{fig:stop-tg10} bend sharply and arises because
  $\chi_1^0 \chi_1^0$-annihilation to $\tau^+\tau^-$ gets significant here.}
\label{fig:stop-tg10-oh2diff}}

The new version of \ds\ offers the possibility to include all squark 
coannihilations in the relic density calculations, but only stop 
coannihilations have proven to affect the result for the mSUGRA framework. 
In this context, the lightest stop is always the lightest squark; its
mass is usually much larger than the LSP mass, unless off-diagonal entries 
in the stop mass matrix become sufficiently large to drive the lightest 
stop mass to small values: this can happen when $|A_0|$ is large.
If the mass of the lightest stop is close to the neutralino mass, 
stop coannihilation effects become important in the neutralino relic 
density calculation~\cite{BDD,Ellisstop,ADS}. In Fig.~\ref{fig:stop-tg10} we 
show the relic density in the $m_{1/2}-m_0$-plane for $\tan \beta=10$, 
$A_0=-2750$ GeV and $\mu>0$. The hatched region is excluded mainly 
due to the neutralino not being the LSP here. The lower right part has 
the $\tilde{\tau}_1$ as the LSP, while $\tilde{t}_1$ is the LSP in most of 
the `bump' to the left. A narrow band close to the lower part of the right 
edge of this `bump', 
i.e.~to the left of the $\Omega{h^2}$ curves, is excluded because of 
the mass limit on $H^0_2$. In other words, the $H^0_2$ mass limit cuts 
away some part of the parameter space with the smallest mass splitting between 
$\tilde{t}_1$ and the LSP\@. In Fig.~\ref{fig:stop-tg10}a we show the relic 
density isolevel curves
without including coannihilation processes. We have also included the 
isolevel curves for the neutralino mass. It is clearly seen that the 
neutralino-neutralino annihilation cross section dramatically increases at 
the $t\bar{t}$ threshold. The relic density then decreases to 
cosmologically interesting values in a band were the mass splitting 
between the $\tilde{t}_1$ and the neutralino is small. 
It is the t-channel exchange of the light stop that boosts the neutralino 
annihilation into $t \bar{t}$. 
In Fig.~\ref{fig:stop-tg10}b we show the results of including coannihilation
processes. Three effects are seen. First of all, the stop coannihilations open
a narrow band of cosmologically interesting density for neutralinos below the
top mass. Secondly, in the almost vertical band above $m(\chi^0_1) = m(t)$,
that was already present in Fig.~\ref{fig:stop-tg10}a, the relic density is
decreased further due to $\tilde{t}_1$ coannihilations.  Finally, we also find
interesting relic densities for high $m_{1/2}$ and low $m_0$ where
$\tilde{\tau}_1$-coannihilations lower the relic density. The same three
features were also found by Ellis, Olive and Santoso, \cite{Ellisstop} Fig.~6d.
Beside the trivial sign difference in the convention used for $A_0$, we also
have chosen the numerical value of $A_0$ slightly lower than done in
\cite{Ellisstop}, in order to obtain a figure as similar to theirs as possible.
The possible explanation for the required change of $|A_0|$ is that we use
different RGE codes, and a given $|A_0|$ therefore results in different low
energy masses. This is not only true for the neutralino and sfermion masses but
also for the Higgs masses. As mentioned, part of the edge of the excluded
region is set by the $H^0_2$ mass limit. The excluded region would be even
larger if we applied the Standard Model Higgs limit ($\sim 114$ GeV) as done in
\cite{Ellisstop} instead of the value $m(H^0_2) \sim 92$ GeV that applies for
$\tan\beta = 10$ \cite{pdg02}.

To see more clearly the effect of coannihilations, in
Fig.~\ref{fig:stop-tg10-oh2diff} we show the difference in relic density,
 $(\Omega_{\chi,\, \rm no\, coann}-\Omega_{\chi,\, \rm coann})/
\Omega_{\chi,\, \rm coann}$, 
versus the neutralino mass along the isolevel curves shown in 
Fig.~\ref{fig:stop-tg10}. We see that in the coannihilation region 
in the upper left corner of Fig.~\ref{fig:stop-tg10}b, 
$\tilde{t}_1$-coannihilations are indeed very important changing 
the relic density by a factor of several 100. When we get to the 
$t \bar{t}$ threshold, the effect of coannihilations decreases 
drastically since the $\chi_1^0 \chi_1^0$ annihilation cross section 
is high.
When the neutralino mass is increased further, the importance of 
coannihilations again increases and above $\sim 275$ GeV the 
$\tilde{\tau}$-coannihilations start getting important.

\section{Conclusions}

We have presented a novel tool for calculating the neutralino relic 
abundances with an estimated precision of 1\% or better, 
assuming masses, widths and couplings of particles in the MSSM
are given. We allow for the most generic coannihilation 
effect in the framework of the MSSM, applying at the same time the state 
of the art technique to trace the freeze-out of a species in the early 
Universe. The code for numerical computations will be 
publicly available in the near future together with a new extended version 
of the \ds\ package \cite{ds}.

As a first example, in this paper we have discussed results valid in the 
mSUGRA setup, the rather restrictive framework most recent analyses of 
coannihilations have focussed on, but, at the same time, general enough to 
illustrate most of the effects that can arise in the MSSM. For a selection of 
mSUGRA models we have given the reader a visual guide into general trends
and their exceptions, such as the tendency of coannihilation effects to lower 
the neutralino relic abundances and the cases in which the opposite can 
happen. We have then performed a broad study of the neutralino relic
density in the mSUGRA parameter space. The features we have spotted are
in agreement with the picture emerging from previous analyses; in particular 
we have discussed the cases of neutralino coannihilations with sleptons (most 
notably stau leptons), with stop squarks, and with charginos. Especially the chargino coannihilations have been treated in more detail than before.

The accuracy of the calculation we performed allowed
us to go into great details in the cases presented. Novel features we
have discussed include, e.g.: (i) a rule of thumb to discriminate the case
when slepton coannihilations are relevant: 25\% mass splitting between
lightest stau and lightest neutralino for a 1\% accuracy on the cosmologically allowed neutralino relic density; (ii) a shift to heavier masses of the cosmological bound on 
the neutralino mass in both the slepton and chargino 
coannihilation cases: in the $\tan \beta=30$ case we studied, we found 565 GeV 
as an upper limit on the neutralino mass from the cosmological bound 
$\Omega_\chi h^2 < 0.2$ in the regime $m_0 \lsim m_{1/2}$, where
slepton coannihilations are important, and 1500 GeV in the regime
$m_0 \gg m_{1/2}$, where chargino coannihilations occur; (iii) a correlation 
between mass splittings, differences in $\Omega_\chi h^2$ and neutralino masses 
in all of the cases presented. 

In an upcoming paper, we will investigate the effects of the 
coannihilation channels in a more general MSSM context.

\section{Acknowledgements}

We thank L.~Bergstr\"om for discussions regarding the calculation of the cross 
section of squarks annihilating into two gluons.
We also thank K.~Matchev for comments.
J.E.\ thanks the Swedish Research Council for support. 
P.U.\ was supported in part by the RTN project under 
grant HPRN-CT-2000-00152 and by the Italian INFN under the
project ``Fisica Astroparticellare''. 
P.U., J.E., and P.G.\ thank the Kavli Institute for Theoretical Physics, 
where parts of this work were completed, for its hospitality.
This research was supported in part by the National Science Foundation under 
grant No.~PHY99-07949.

\clearpage
\appendix
\section{Included coannihilations}
\label{sec:processes}

Here we  list all $2 \rightarrow 2$ tree-level coannihilation processes 
with sfermions, charginos and neutralinos.  
All the processes are included in the \ds\ code, but not all of them have been
used in the calculation we report on. As mentioned earlier, we have found that
an accuracy of 1\% on the relic neutralino density in the cosmologically
interesting interval is obtained by including coannihilations for particles
lighter than $1.5m_\chi$. Consequently, some sparticles never get included in
the initial state. For the sparticles that satisfy the mass difference
criterium, we have included all coannihilation processes and for each of these,
all the exchange channels. Furthermore, the \ds\ relic density
code always includes the one loop neutralino annihilation into $\gamma\gamma$,
$\gamma{Z}$ and $gg$.

It should be noted that we have not included all flavour-changing 
charged current diagrams. The \ds\ vertex code for the charged current 
couplings is written in a general form that includes all possible 
flavour-changing (and flavour-conserving) vertices. The 
flavour-conserving couplings are much larger than the flavour-changing. 
For the sfermion coannihilations with charged currents we only take the 
flavour-conserving contributions, while for the chargino coannihilations 
we include the flavour-changing contributions as well. In a future 
version of \ds, we may as well include the flavour-changing processes 
for the sfermion coannihilations, even if they are not expected to
be important.

We have used the notation $\ft$ for sfermions and $f$ for
fermions. Whenever the isospin of the sfermion/fermion is important,
it is indicated by an index $u$ ($T_3 = 1/2$) or $d$ ($T_3 = -1/2$).
The sfermions have an additional mass eigenstate index, that can take
the values 1 and 2 (except for the sneutrinos which only have one mass
eigenstate).  A further complication to the notation is when the
sfermions and fermions in initial, final and exchange state can belong
to different families. Primes will be used to indicate when we have
this freedom to choose the flavour. So, e.g.~$\ft_u$ and $f_u$ will
belong to the same family while $\ft_u$ and ${f}^{\prime}_u$ can belong to the
same or to different families. Note that the colour index of (s)quarks
as well as gluons ($g$) and gluinos ($\tilde{g}$) is suppressed.

Besides the sfermions we also have neutralinos and charginos in the
initial states. The notation used for these are the following. The
neutralinos are denoted by $\chi^0_j$ with the index running from 1 to
4.  The charginos are similarly denoted $\chi^\pm_j$ with the index
taking the values 1 and 2.

In the table in this appendix, a common notation is introduced for
gauge and Higgs bosons in the final state. We denote these with $B$
with an upper index indicating the electric charge. So $B^0$ means
$H^0_1, H^0_2, H^0_3, Z, \gamma$ and $g$ while $B^\pm$ is $H^\pm$ and
$W^\pm$.  We will use additional lower indices $m$ and $n$ when we
have more than one boson in the final state. Thus indicating that the
bosons can be either different or identical. Note that the case of two
different bosons also includes final states with one gauge boson and
one higgs boson.

The table has been made very general. This means that when a set of
initial and final state (s)particles have been specified, the given
process might not run through all the exchange channels listed for the
generic process. Exceptions occur whenever an exchange (s)particle
does not couple to the specific choice of initial and/or final
state. As an example we see that since the photon does not couple to
neutral (s)particles, none of the exchange channels listed for the
generic process $\ft_i + \chi^0_j \into B^0 + f$ actually exist for
the specific process $\snu + \chi^0 \into \gamma + \nu$. All these
exceptions can be found in the extended tables in Ref.\
\cite{ds-manual}. Also note that the list of processes is not complete
with respect to trivial charge conjugation. For each process of
nonvanishing total electric charge in the initial state there exist
another process which is obtained by charge conjugation.


\TABLE[h]{
\begin{tabular}{p{4cm}p{3cm}p{2cm}p{2.0cm}p{0.5cm}}
 & \multicolumn{4}{c}{Diagrams} \\ \cline{2-5}
Process & s  & t & u & p \\
\hline
~ & \\[-2.5ex]
$\chi^0_i \chi^0_j \into B^0_m B^0_n$ & $H^0_{1,2,3},Z$ & $\chi^0_k$ & $\chi^0_l$ \\
$\chi^0_i \chi^0_j \into B^-_m B^+_n$ &Ê$H^0_{1,2,3},Z$ & $\chi^+_k$ & $\chi^+_l$ \\
$\chi^0_i \chi^0_j \into f \bar{f}$ &Ê$H^0_{1,2,3},Z$ & $\ft_{1,2}$ & $\ft_{1,2}$ \\[1ex]
\hline
~ & \\[-2.5ex]
$\chi^+_i \chi^0_j \into B^+_m B^0_n$ & $H^+,W^+$ & $\chi^0_k$ & $\chi^+_l$ \\
$\chi^+_i \chi^0_j \into f_u \bar{f}_d$ &Ê$H^+,W^+$ & $\ft^\prime_{d_{1,2}}$ & $\ft^\prime_{u_{1,2}}$ \\[1ex] \hline
~ & \\[-2.5ex]
$\chi^+_i \chi^-_j \into B^0_m B^0_n$ & $H^0_{1,2,3},Z$ & $\chi^+_k$ & $\chi^+_l$ \\
$\chi^+_i \chi^-_j \into B^+_m B^-_n$ & $H^0_{1,2,3},Z,\gamma$ & $\chi^0_k$ & {} \\
$\chi^+_i \chi^-_j \into f_u \bar{f}_u$ &Ê$H^0_{1,2,3},Z,\gamma$ & $\ft^\prime_{d_{1,2}}$ & {} \\
$\chi^+_i \chi^-_j \into \bar{f}_d f_d$ &Ê$H^0_{1,2,3},Z,\gamma$ & $\ft^\prime_{u_{1,2}}$ & {} \\
$\chi^+_i \chi^+_j \into B^+_m B^+_n$ & {} & $\chi^0_k$ & $\chi^0_l$ \\[1ex]
\hline
~ & \\[-2.5ex]
$\ft_i \chi^0_j     \into B^0 f$   & $f$   & $\ft_{1,2}$ & $\chi^0_l$ \\
$\ft_{d_i} \chi^0_j \into B^- f_u$ & $f_d$ & $\ft_{u_{1,2}}$ & $\chi^+_l$    \\
$\ft_{u_i} \chi^0_j \into B^+ f_d$ & $f_u$ & $\ft_{d_{1,2}}$ & $\chi^+_l$ \\[1ex]
\hline
~ & \\[-2.5ex]
$\ft_{d_i} \chi^+_j \into B^0 f_u$ & $f_u$ & $\ft_{d_{1,2}}$ & $\chi^+_l$    \\
$\ft_{u_i} \chi^+_j \into B^+ f_u$ & {}    & $\ft_{d_{1,2}}$ & $\chi^0_l$    \\
$\ft_{d_i} \chi^+_j \into B^+ f_d$ & $f_u$ & {}              & $\chi^0_l$    \\
$\ft_{u_i} \chi^-_j \into B^0 f_d$ & $f_d$ & $\ft_{u_{1,2}}$ & $\chi^+_l$ \\
$\ft_{u_i} \chi^-_j \into B^- f_u$ & $f_d$ & {}              & $\chi^0_l$ \\
$\ft_{d_i} \chi^-_j \into B^- f_d$ &       & $\ft_{u_{1,2}}$ & $\chi^0_l$ \\[1ex]
\hline
~ & \\[-2.5ex]
$\ft_{d_i} \ft_{d_j}^\ast \into B^0_m B^0_n$ & $H^0_{1,2,3},Z,g$ & $\ft_{d_{1,2}}$    &
$\ft_{d_{1,2}}$ & p \\
$\ft_{d_i} \ft_{d_j}^\ast \into B^-_m B^+_n$ & $H^0_{1,2,3},Z,\gamma$ & $\ft_{u_{1,2}}$ &
{}              & p \\
$\ft_{d_i}\ft_{d_j}^{\prime\ast}\into f^{\prime\prime}_d \bar{f}^{\prime\prime\prime}_d$ & $H^0_{1,2,3},Z,\gamma,g$ & $\chi^0_k,\gt$ &
{}              & {}  \\
$\ft_{d_i} \ft_{d_j}^{\prime\ast} \into f^{\prime\prime}_u \bar{f}^{\prime\prime\prime}_u$ & 
$H^0_{1,2,3},Z,\gamma,g$ & $\chi^+_k$ & 
{}              & {}  \\
$\ft_{d_i}{\ft}^\prime_{d_j} \into f_d {f}^\prime_d$ & {} & $\chi^0_k,\gt$  & 
$\chi^0_l,\gt$& {} \\[1ex]
\hline
~ & \\[-2.5ex]
$\ft_{u_i}\ft_{d_j}^\ast\into B^+_mB^0_n$& $H^+,W^+$ & $\ft_{d_{1,2}}$ & $\ft_{u_{1,2}}$ & p \\
$\ft_{u_i}\ft_{d_j}^{\prime\ast}\into f^{\prime\prime}_u \bar{f}^{\prime\prime\prime}_d$ & $H^+,W^+$ & $\chi^0_k,\gt$ & {} & {}\\
$\ft_{u_i}{\ft'}_{d_j}\into f^{\prime\prime}_u f^{\prime\prime\prime}_d$ & {} & $\chi^0_k,\gt$ & $\chi^+_l$ & {} \\[1ex]
\hline
\end{tabular}
\caption{Included coannihilation processes through $s-$, $t-$, $u-$channels 
and four-point interactions (p). For the $\ft_{d_i} \ft_{d_j}^{(\ast)}$ processes
the corresponding process for up-type sfermions can be obtained by interchanging
the $u$ and $d$ indices.}
\label{tab:coanns}
}


\section{A note about internal degrees of freedom}
\label{appdof}

Here we describe a technical detail in the calculation, which we find
useful 
to specify.
If we look at Eqs.~(\ref{eq:weff}) and (\ref{eq:sigmavefffin2}) we see that we
have a freedom on how to treat particles degenerate in mass.  For example, the
charginos, $\chi^\pm$, (where the mass eigenstate index is implicit) can be
treated either as (a) two separate species $\chi^+$ and $\chi^-$, each with
internal degrees of freedom $g_{\chi^+}=g_{\chi^-}=2$, or (b) a single species
$\chi^\pm$ with $g_{\chi^\pm}=4$ internal degrees of freedom.  Of course the
two views are equivalent, we just have to be careful to include the $g_{i}$'s
consistently whichever view we take.  In a), we have the advantage that all the
$W_{ij}$ that enter into Eq.~(\ref{eq:weff}) enter as they are, i.e.\ without
any correction factors for the degrees of freedom. On the other hand we get
many terms in the sum that are identical and we need some book-keeping
machinery to avoid calculating identical terms more than once. On the other
hand, with option b), the sum over $W_{ij}$ in Eq.~(\ref{eq:weff}) is much
simpler only containing terms that are not identical (except for the trivial
identity $W_{ij}=W_{ji}$ which is easily taken care of).  However, the
individual $W_{ij}$ will be some linear combinations of the more basic $W_{ij}$
entering in option a), where the coefficients have to be calculated for each
specific type of initial condition.

We have chosen to work with option b) since this most easily gives an
efficient numerical code. Denoting the $W_{ij}$'s in option b) with a
prime, we can derive \cite{ds-manual} the following relations between
the two different views,

\begin{equation}
  \left\{ \begin{array}{lcl}
  W'_{\chi_{i}^0 \chi_{j}^\pm} & \equiv & W_{\chi_{i}^0 \chi_{j}^+} = 
    W_{\chi_{i}^0 \chi_{j}^-} \quad , \quad \forall\ i=1,\ldots,4,\ 
    j=1,2 \anl
  W'_{\chi_{i}^\pm \chi_{j}^\pm} & \equiv & \frac{1}{2} 
  \left[ W_{\chi_{i}^+ \chi_{j}^+} +  
  W_{\chi_{i}^+ \chi_{j}^-}\right] = 
  \frac{1}{2} \left[ W_{\chi_{i}^- \chi_{j}^-} +  
  W_{\chi_{i}^- \chi_{j}^+}\right] \quad , \quad \forall\ i=1,2,\ j=1,2 \anl
  W'_{\chi_{i}^0 \tilde{f}^\prime_{k}} & \equiv & W_{\chi_{i}^0 \tilde{f}_{k}}
  \quad , \quad \forall i=1,\ldots 4,\ k=1,2 \anl
  W'_{\chi_{c}^\pm \tilde{f}^\prime_{k}} & \equiv &
    \frac{1}{2} \left[ 
    W_{\chi_{c}^+ \tilde{f}_{k}} + 
    W_{\chi_{c}^+ \tilde{f}_{k}^*}
    \right]
  \quad , \quad \forall c=1,2,\ k=1,2 \anl
  W'_{\tilde{f}^\prime_{k} \tilde{f}^\prime_{l}} & \equiv & 
    \frac{1}{2} \left[
    W_{\tilde{f}_{k} \tilde{f}_{l}} + 
    W_{\tilde{f}_{k} \tilde{f}_{l}^*} \right] 
  \quad , \quad \forall k=1,2,\ l=1,2 \anl
  \end{array} \right.
\end{equation}
Where $\tilde{f}$ denotes a generic sfermion. We have neglected in this listing the 
trivial colour averaging for squarks, i.e. $W'_{\tilde{q}^\prime_{k} \tilde{q}^\prime_{l}} 
\equiv  \frac{1}{2} \frac{1}{9}\sum_{a,b=1}^3 \left[W_{\tilde{q}_{k}^a \tilde{q}_{l}^b} + 
W_{\tilde{q}_{k}^a \tilde{q}_{l}^{b*}} \right]$ and similarly for the other cases.

\clearpage

\section{Example models}
\label{appexmodels}

Here we list the parameters and some properties of the example models
considered in some of the figures.

\TABLE[h]{
\centering
\begin{tabular}{lrrrrrr}
Model        & A & B &
C & D & E & F\\ \hline
$m_0$        & 387.0     & 898.0  & 1427.0    & 160.0 & 76.7 & 193.3 \\
$m_{1/2}$    & 950.0     & 1496.0 & 320.0     & 780.0 & 348.8 & 882.1 \\
$\tan \beta$ & 10.0      & 50.2   & 10.0      & 5.0 & 10.0 & 10.0 \\
$A_0$        & $-3770.0$ & 999.0  & $-2750.0$ & 0.0 & 0.0 & 0.0 \\
$sign(\mu)$  & $+$       & $+$    & $+$       & $+$ & $+$ & $+$ \\  \hline
$m_{\chi_1^0}$       & 406.4 & 646.5  & 133.8  & 325.8 & 138.5 & 371.1  \\
$m_{\chi_2^+}$       & 757.5 & 1184.4 & 256.1  & 603.6 & 255.0 & 686.9 \\
$m_{\tilde{e}_2}$    & 523.7 & 1050.6 & 1430.0 & 331.1 & 156.2 & 379.7 \\
$m_{\tilde{\mu}_2}$  & 523.7 & 1050.6 & 1430.0 & 331.1 & 156.2 & 379.7  \\
$m_{\tilde{\tau}_1}$ & 407.4 & 665.1 & 1396.0 & 329.1  & 148.0 & 371.8  \\
$m_{\tilde{t}_1}$    & 578.7 & 2297.3 & 200.0  & 1176.0 & 542.4 & 1331.8 \\
\hline
$\Omega_\chi h^2|_{\mbox{w. coanns}}$  & 0.135 & 0.155 & 19.7 & 0.109 &
0.115
& 0.115 \\
$\Omega_\chi h^2|_{\mbox{w/o coanns}}$ & 1.43  & 0.137 & 73.2 & 0.981 &
0.230
& 1.26 \\ \hline
Shown in Fig. & \ref{fig:weff-sf-coann} & \ref{fig:weff-sf-incr} &
\ref{fig:weff-stop} & \ref{fig:weff-smuimp} & \ref{fig:slweff}a &
\ref{fig:slweff}b \\ \hline
\end{tabular}
\caption{Model parameters and some properties of the example models
discussed in the text and figures. The sparticle masses are calculated with
ISASUGRA 7.64.}
\label{tab:examples}
}
   


\end{document}